\documentclass[12pt]{article}

\usepackage[T1]{fontenc}
\usepackage[utf8]{inputenc}

\usepackage{amsmath,amssymb,amsthm,mathtools}
\usepackage{bm}
\usepackage{subcaption}
\usepackage{graphicx}
\usepackage{array,booktabs,tabularx,multirow}
\usepackage{pdflscape}
\usepackage{enumitem}
\usepackage{appendix}
\usepackage{newunicodechar}
\newunicodechar{−}{-}
\usepackage{tikz}
\usetikzlibrary{matrix,positioning}
\usepackage{natbib}
\usepackage{setspace}
\usepackage{tocloft}

\usepackage[dvipsnames]{xcolor}

\usepackage[
    colorlinks = true,
    linkcolor  = blue, 
    citecolor  = blue,  
    urlcolor   = blue      
]{hyperref}

\allowdisplaybreaks

\DeclareUnicodeCharacter{2234}{\therefore}
\DeclareUnicodeCharacter{202F}{~}

\newcolumntype{P}[1]{>{\raggedright\arraybackslash}p{#1}}
\newcolumntype{Y}{>{\raggedright\arraybackslash}X}

\usepackage[margin=1in]{geometry}

\usepackage{authblk}

\cftpagenumbersoff{figure}
\cftpagenumbersoff{table}

\newcommand{\keywords}[1]{%
  \vspace{0.5em}%
  \noindent\textbf{Keywords: }#1
}

\title{Evolution under Stochastic Transmission:\\ Mutation-Rate Modifiers}

\author[1]{Elisa Heinrich-Mora}
\author[1]{Marcus Feldman\thanks{Corresponding author: mfeldman@stanford.edu}}
\affil[1]{Department of Biology, Stanford University, Stanford, CA, USA}

\begin{document}
\begin{titlepage}
    \thispagestyle{empty}
    \centering
     \vspace*{4em}

    {\LARGE \bfseries Evolution under Stochastic Transmission:\\
    Mutation-Rate Modifiers\par}
    \vspace{1em}
    {\large Elisa Heinrich-Mora, Marcus Feldman\footnote{Corresponding author: mfeldman@stanford.edu}\par}
    \vspace{1em}
    {\small Department of Biology, Stanford University, Stanford, CA, USA\par}
     \vspace*{\fill}

\end{titlepage}

\begin{abstract}
\noindent Evolutionary analyses of large populations commonly incorporate stochasticity through temporal variation in selection while treating genetic transmission as fixed. Much less attention has been given to stochasticity in transmission itself. We study a selected locus with alleles \(A\) and \(a\) under constant selection, linked to a neutral modifier locus whose alleles $M_1$ and $M_2$ control the mutation rate from \(A\) to \(a\). Under constant transmission, the Reduction Principle applies: near a mutation--selection balance where $M_1$ is fixed with mutation rate $u_1$, a rare allele $M_2$ invades if  its associated rate $u_2$ is smaller than $u_1$, but cannot invade if $u_2$ is larger than $u_1$. This result holds for both haploid and diploid populations and is independent of recombination, which affects only the rate, not the direction, of evolutionary change. We extend this framework by allowing the mutation rate associated with the invading modifier to fluctuate randomly across generations. In this stochastic setting, invasion is no longer determined by mean mutation rates alone. Instead, it depends on the temporal distribution of mutation rates, the strength of selection at the selected locus, and the recombination rate between modifier and target. Stochastic transmission and recombination therefore do not merely rescale deterministic predictions based on the Reduction Principle; they can alter the direction of selection on modifier alleles.
\end{abstract}

\newpage
\doublespacing
\section*{Introduction}
\noindent Why do genetic systems have the structure they do?  Evolutionary theory has been successful at explaining changes in allele frequencies under selection, yet it has often treated the rules of inheritance---mutation, recombination, and the structure of gene action---as fixed features of the system. This view implies that selection acts on variation without altering the processes that produce it. Yet empirical evidence and theory both show that transmission parameters can themselves be evolvable traits. Evolutionary theory must therefore account not only for changes in genetic variation, but for the evolution of the genetic architecture that produces that variation.

\noindent \emph{Modifier theory} provides a formal framework for addressing this problem.  It focuses on loci whose alleles alter the transmission or genetic structure of other loci under selection---including modifiers of mutation rate, recombination, dominance, or epistasis---and asks whether the frequency of a rare modifier allele increases when it is introduced into a population near equilibrium.  For modifiers of transmission parameters, such as mutation rate, that have no direct effect on viability or fertility; selection arises solely through the statistical associations the modifier generates with selected genetic backgrounds.  Invasion analysis makes this indirect selection explicit and embeds the evolution of genetic architecture within standard population--genetic theory \citep{altenberg1984}.

\noindent A central result of modifier theory is the \emph{Reduction Principle}.  Under deterministic assumptions ---infinite population size, random mating, constant viability selection, and fixed transmission parameters--- a rare modifier allele that \emph{reduces} the mutation rate at loci maintained at mutation--selection balance, or that \emph{reduces} recombination between epistatically interacting loci that generate linkage disequilibrium, increases in frequency. Conversely, a modifier allele that increases mutation, recombination or migration rates cannot invade and is eliminated when rare \citep{feldman1972selection,karlin1974towards,teague1976result,balkau1973selection,feldman1976krakauer,liberman1986modifiers,feldman1980evolution,feldman1986evolutionary,feldman1996population,altenberg1984,altenberg2017unified}.  The force of the Reduction Principle lies in its generality: it follows from the structure of indirect selection under deterministic dynamics, rather than from the details of any particular genetic system.  Subsequent work has therefore focused on identifying conditions under which its assumptions fail and on determining which departures from them---such as finite population size, nonrandom mating, or temporal variation in selection---alter its conclusions.

\noindent There is substantial literature on the effect of temporal variation in \emph{selection} itself.  In these models, fitnesses fluctuate across generations and evolutionary change is governed by long--term multiplicative growth. Gillespie \citep{gillespie1972,gillespie1973}, Cook and Hartl \citep{cookHartl1974}, and Karlin and Liberman \citep{karlinLiberman1975} showed that fluctuating selection can maintain polymorphism and favor the allele with the highest geometric mean fitness, a quantity shaped by higher moments of the fitness distribution as well as its mean. Closely related results from population growth theory show that, in randomly varying environments, long--term growth is governed by geometric rather than arithmetic averages \citep{lewontin1969population}. 

\noindent Modifier theory has also been extended to variable selective environments while retaining its core logic of indirect selection.  When selection coefficients vary in space or time, changing fitness landscapes generate indirect selection on modifier alleles \citep{charlesworth1976recombination,carja2014evolution}.  Environmental variability can favor nonzero mutation, recombination, or phenotypic switching rates by increasing the geometric mean fitness across environmental cycles, even when such rates are disfavored under constant conditions \citep{liberman2011evolution}. Related results show that, in periodically varying environments, unconditional dispersal can be favored for the same reason: modifiers that alter transmission or movement affect long--term growth through their impact on multiplicative fitness across environmental states \citep{schreiber2011evolution}.  Simulation studies of multigenic mutation--rate modifiers in sexual populations further show that the \emph{form} of selection matters: stabilizing versus directional selection can differentially bias the evolution of mutation rates by altering the statistical associations between modifier alleles and the phenotypic tails of the trait distribution \citep{rozhok2021silico}.  Together, these results show that the evolution of transmission parameters such as mutation rate should depend not only on the magnitude of environmental change, but on its temporal structure and distributional properties.

\noindent  Empirical studies across taxa also motivate relaxing the assumption of fixed transmission.  Comparative and mutation--accumulation studies document substantial variation in mutation rates arising from physiological state, genomic context, and external conditions, often with little evidence of tight coupling to organismal fitness \citep{drake_rates_1998,baer_mutation_2007,lynch_evolution_2010}. Environmental factors such as temperature, nutrient limitation or starvation, oxidative or respiratory environment, desiccation--rehydration cycles, and exposure to mutagens, including ultraviolet radiation, can alter mutation processes in a manner that acts broadly across individuals \citep{galhardo_mutation_2007,foster_stress-induced_2007}.  In such cases, the mutation rate is approximately \emph{uniform across the population within a generation}, while varying across generations as conditions change.  This temporal variation contrasts with models in which mutation rates differ among individuals within the same generation due to intrinsic molecular noise or lineage-specific effects \citep{ninio_transient_1991}.

\noindent Several recent theoretical studies have explored related but distinct departures from fixed transmission. Lobi\'nska et al.\ \citep{lobinska_phenotype_2023} analyzed populations with discrete mutation-rate phenotypes connected by switching, focusing on rates of adaptation rather than modifier invasion. Other work has considered deterministic heterogeneity in mutation rates that is directly coupled to individual fitness, so that mutation rates vary among individuals within a generation but are not stochastic in time. In this case, transmission covaries with fitness and mutation modification is no longer neutral, so the assumptions underlying the Reduction Principle are violated \citep{ram_evolution_2012}. Weissman \citep{weissman_stress-induced_nodate} examined the evolutionary consequences of variable mutation rates in a constant selective environment, emphasizing how relaxing fixed transmission alters long-term dynamics.

\noindent Motivated by these considerations, we study models in which stochasticity enters through the \emph{transmission} process rather than through selection.  Selection is held constant, while the mutation rate is uniform across the population within each generation but fluctuates randomly across generations. We analyze the invasion of a rare modifier allele that alters the average mutation rate and may introduce inter-generational variance in that rate.  The central question is whether indirect selection on the modifier allele continues to conform to the Reduction Principle when transmission itself is stochastic.

\noindent Formally, let $\{T_t\}_{t\ge 0}$ denote a sequence of transmission parameters, where $T_t$ is the population--wide value in generation $t$.  The $T_t$ are independent and identically distributed with mean $\mu=\mathbb{E}[T_t]$ and variance $\sigma^2=\mathrm{Var}(T_t)$, supported on a biologically realistic interval; no particular distributional form is assumed. Classical modifier theory corresponds to the case $\sigma^2=0$, in which $T_t\equiv\mu$ for all $t$. When transmission fluctuates through time, modifier alleles with identical mean effects can nevertheless differ in invasibility, because higher--order moments of the transmission process affect the long--term multiplicative growth of rare modifier allele.  This growth is governed by the top Lyapunov exponent of the linearized recursion for modifier haplotype frequencies, which depends on the full distribution of $\{T_t\}$.  Transmission variability is therefore not extraneous noise, but a determinant of evolutionary outcome.  Extending modifier theory to temporally fluctuating, population--wide transmission parameters thus provides a natural framework for analyzing how endogenous variability in inheritance shapes the evolution of genetic systems.

\section{Model Set-up} \label{sec: model_setup_unified}
\noindent A large population is considered to have two biallelic loci: a selected locus ($A/a$) under constant selection and a linked, selectively neutral modifier ($M_1/M_2$) that controls the \emph{forward} mutation rate $A\to a$. Back mutation $a\to A$ is absent. The recombination fraction between loci is $R\in[0,\tfrac12]$, constant through time. The life cycle within each generation is
\[
\mathbf x \xrightarrow{\ \text{selection}\ }\mathbf x'
\xrightarrow{\ \text{recombination}\ }\mathbf x''
\xrightarrow{\ \text{mutation}\ }\mathbf x^{(t+1)}.
\]
\noindent Let $\mathbf x=(x_1,x_2,x_3,x_4)$ denote haplotype frequencies $(AM_1,aM_1,AM_2,aM_2)$ with $\sum_i x_i=1$.  After selection and recombination the frequencies are $\mathbf x'$ and $\mathbf x''$, respectively; $\mathbf x^{(t+1)}$ denotes the state at the start of generation $t{+}1$.

\paragraph{Selection.} Selection acts only at the $A/a$ locus.

\noindent In haploids, viabilities are $W_A=1$ and $W_a=1-s$ with $s\in(0,1]$, yielding mean fitness
\[
\bar W_{\mathrm{hap}}=(x_1+x_3)+(1-s)(x_2+x_4),
\]
and post-selection frequencies
\[
x_1'=\frac{x_1}{\bar W_{\mathrm{hap}}},\quad
x_2'=\frac{(1-s)x_2}{\bar W_{\mathrm{hap}}},\quad
x_3'=\frac{x_3}{\bar W_{\mathrm{hap}}},\quad
x_4'=\frac{(1-s)x_4}{\bar W_{\mathrm{hap}}}.
\]

\noindent In diploids, viabilities at the selected locus are additive:
\[
W_{11}=1\ (AA),\quad W_{12}=1-s\ (Aa),\quad W_{22}=1-2s\ (aa),
\]
with $s\in(0,1)$.  Under random mating the mean fitness is
\[
\bar W_{\mathrm{dip}}=(x_1+x_3)^2 W_{11}
+2(x_1+x_3)(x_2+x_4)W_{12}
+(x_2+x_4)^2 W_{22}.
\]
\noindent The resulting post-selection gamete frequencies $\mathbf x'=\mathbf x'(\mathbf x;s)$ are the standard two-locus frequencies with viabilities $W_{11},W_{12},W_{22}$.

\paragraph{Recombination.} Recombination acts on $\mathbf x'$ in haploids and diploids at the gamete level. Let
\[
D' = x_1' x_4' - x_2' x_3'
\]
denote linkage disequilibrium after selection.  After recombination,
\[
x_1''=x_1'-R D',\quad
x_2''=x_2'+R D',\quad
x_3''=x_3'+R D',\quad
x_4''=x_4'-R D'.
\]

\paragraph{Mutation.} Forward mutation $A\to a$ acts on $A$-bearing gametes after recombination and depends on the modifier genotype.

\noindent In haploids, modifier allele $M_1$ induces a constant mutation rate $u_1\in[0,1)$, while modifier allele $M_2$ induces a \emph{population-wide, time-dependent} mutation rate $u_{2,t}\in[0,1)$ that is identical for all $M_2$ carriers within generation $t$.  The mutation step is
\begin{equation}\label{eq:hap_mutation}
\begin{aligned}
x_1^{(t+1)}&=(1-u_1)\,x_1'',&
x_2^{(t+1)}&=x_2''+u_1\,x_1'',\\
x_3^{(t+1)}&=(1-u_{2,t})\,x_3'',&
x_4^{(t+1)}&=x_4''+u_{2,t}\,x_3''.
\end{aligned}
\end{equation}

\noindent In diploids, the mutation rate depends on the modifier genotype of the parent.  Gametes produced by $M_1M_1$ individuals mutate at a constant rate $u_1$.  Gametes produced by $M_1M_2$ individuals mutate at a population--wide, time--dependent rate $u_{2,t}$ in generation $t$, which is identical for all such individuals.  Gametes produced by $M_2M_2$ individuals mutate at rate $u_3$.  Thus, mutation rates may differ among modifier genotypes, but for each genotype the mutation rate is uniform across the population within a generation; only the rate associated with heterozygotes varies across generations.

\paragraph{Full recursion.} Substituting the above steps yields the evolutionary recursions.
\noindent\emph{Haploids:}
\begin{equation}\label{eq:hap_recursion}
\begin{aligned}
x_1^{(t+1)} &= \frac{x_1}{\bar W_{\mathrm{hap}}}(1-u_1)
 - \frac{R(1-s)(x_1 x_4-x_2 x_3)}{\bar W_{\mathrm{hap}}^2}(1-u_1),\\
x_2^{(t+1)} &= \frac{(1-s)x_2}{\bar W_{\mathrm{hap}}}
 + \frac{R(1-s)(x_1 x_4-x_2 x_3)}{\bar W_{\mathrm{hap}}^2}
 + \frac{x_1}{\bar W_{\mathrm{hap}}}u_1
 - \frac{R(1-s)(x_1 x_4-x_2 x_3)}{\bar W_{\mathrm{hap}}^2}u_1,\\
x_3^{(t+1)} &= \frac{x_3}{\bar W_{\mathrm{hap}}}(1-u_{2,t})
 + \frac{R(1-s)(x_1 x_4-x_2 x_3)}{\bar W_{\mathrm{hap}}^2}(1-u_{2,t}),\\
x_4^{(t+1)} &= \frac{(1-s)x_4}{\bar W_{\mathrm{hap}}}
 - \frac{R(1-s)(x_1 x_4-x_2 x_3)}{\bar W_{\mathrm{hap}}^2}
 + \frac{x_3}{\bar W_{\mathrm{hap}}}u_{2,t}
 + \frac{R(1-s)(x_1 x_4-x_2 x_3)}{\bar W_{\mathrm{hap}}^2}u_{2,t}.
\end{aligned}
\end{equation}  with  $\bar W_{\mathrm{hap}}=(x_1+x_3)+(1-s)(x_2+x_4)$.

\noindent\emph{Diploids} \citep{TWOMEY1990320}:
\begin{equation}\label{eq:dip_recursion}
\begin{aligned}
\bar W_{\mathrm{dip}}\,x_1^{(t+1)} &=
(1-u_1)(x_1^2 W_{11}+x_1 x_2 W_{12})
+(1-u_{2,t})\big[x_1 x_3 W_{11}+x_1 x_4 W_{12}-R W_{12}(x_1 x_4-x_3 x_2)\big],\\
\bar W_{\mathrm{dip}}\,x_2^{(t+1)} &=
x_2^2 W_{22}+x_1 x_2 W_{12}+x_3 x_2 W_{12}+x_2 x_4 W_{22}
+R W_{12}(x_1 x_4-x_3 x_2)\\
&\quad
+u_1(x_1^2 W_{11}+x_1 x_2 W_{12})
+u_{2,t}\big[x_1 x_3 W_{11}+x_1 x_4 W_{12}-R W_{12}(x_1 x_4-x_3 x_2)\big],\\
\bar W_{\mathrm{dip}}\,x_3^{(t+1)} &=
(1-u_{2,t})\big[x_3^2 W_{11}+x_3 x_4 W_{12}
+x_1 x_3 W_{11}+x_3 x_2 W_{12}+R W_{12}(x_1 x_4-x_3 x_2)\big],\\
\bar W_{\mathrm{dip}}\,x_4^{(t+1)} &=
x_4^2 W_{22}+x_1 x_4 W_{12}+x_3 x_4 W_{12}+x_2 x_4 W_{22}
-R W_{12}(x_1 x_4-x_3 x_2)\\
&\quad
+u_{2,t}\big[x_3^2 W_{11}+x_3 x_4 W_{12}
+x_1 x_3 W_{11}+x_3 x_2 W_{12}+R W_{12}(x_1 x_4-x_3 x_2)\big].
\end{aligned}
\end{equation}
 with $\bar W_{\mathrm{dip}}=(x_1+x_3)^2 W_{11}+2(x_1+x_3)(x_2+x_4)W_{12}+(x_2+x_4)^2 W_{22}$.

\noindent Recursion systems \eqref{eq:hap_recursion} and \eqref{eq:dip_recursion} form the basis for the invasion analysis in which $M_2$ is initially rare and the resident $M_1$ is close to its mutation–selection equilibrium.

\section{Invasion Analysis of Modifier Allele $M_2$}\label{sec:invasion_unified}
\noindent We ask whether a rare modifier allele $M_2$ increases in frequency when introduced into a population fixed for $M_1$.  Let $\mathbf x_t=(x_{1,t},x_{2,t},x_{3,t},x_{4,t})^\top$ denote the haplotype frequencies of $AM_1$, $aM_1$, $AM_2$, and $aM_2$ at generation $t$.  When $M_2$ is absent, the system reduces to a one--locus mutation--selection model at $A/a$ with equilibrium
\[
\hat{\mathbf x}=(\hat x_1,\hat x_2,0,0)^\top,
\]
where $\hat x_1$ and $\hat x_2$ denote the resident mutation--selection balance under mutation rate $u_1$.

\noindent We restrict attention to parameter values for which this equilibrium is polymorphic, ensuring a well-defined mutation--selection balance from which invasion can be analyzed.  Specifically,
\begin{subequations}\label{eq:mut_sel_balance}
\begin{align}
\text{haploids:}\quad
(\hat x_1,\hat x_2)
&=\Big(\tfrac{s-u_1}{s},\,\tfrac{u_1}{s}\Big),
\qquad 
\hat{\bar W}_{\mathrm{hap}}=1-u_1,
\qquad 0<u_1<s,
\label{eq:4a}\\[4pt]
\text{diploids:}\quad
(\hat x_1,\hat x_2)
&=\Big(\tfrac{s-u_1+su_1}{s(1+u_1)},\,\tfrac{u_1}{s(1+u_1)}\Big),
\qquad 
\hat{\bar W}_{\mathrm{dip}}=\tfrac{1-u_1}{1+u_1},
\qquad 0<u_1<\tfrac{s}{1-s}.
\label{eq:4b}
\end{align}
\end{subequations}
\noindent These conditions ensure that mutation maintains the allele $a$ against selection, producing a stable resident equilibrium. The conditions in \eqref{eq:mut_sel_balance} therefore ensure a stationary, internally stable genetic background near which the growth rate of a rare mutant can be meaningfully linearized.

\noindent Linearizing the full two--locus recursion at $\hat{\mathbf x}$ yields the Jacobian
\[
\mathbf J=
\begin{pmatrix}
\mathbf P & \mathbf C\\[3pt]
\mathbf 0 & \mathbf F
\end{pmatrix},
\]
where $\mathbf P$ governs the internal stability of the resident $(AM_1,aM_1)$ subsystem, $\mathbf C$ captures first--order coupling from resident to rare haplotypes, and $\mathbf F$ governs the external stability of the equilibrium with respect to invasion by $M_2$.  Because $\mathbf J$ is block--triangular, the invasion criterion depends solely on $\mathbf F$.

\noindent Let $\bm v_t=(x_{3,t},x_{4,t})^\top$ denote the frequencies of the rare $M_2$--bearing haplotypes. Near the resident mutation--selection balance,
\[
\bm v_{t+1}=\mathbf F\,\bm v_t,
\]
and $M_2$ invades if the Perron--Frobenius eigenvalue (leading eigenvalue) $\rho(\mathbf F)$ exceeds unity.

\noindent Substituting $(\hat x_1,\hat x_2)$ from \eqref{eq:mut_sel_balance} into the $x_{3,t+1}$ and $x_{4,t+1}$ recursions in \eqref{eq:hap_recursion} and \eqref{eq:dip_recursion}, and retaining terms to first order in the rare haplotype frequencies neglecting terms of order $x_{3,t}^2$, $x_{4,t}^2$, and higher), yields the following invasion matrices.

\noindent For haploids: 
\begin{equation}\label{eq:F_hap}
\mathbf F_{\mathrm{hap}}
=\frac{1}{1-u_1}
\begin{pmatrix}
(1-u_{2,t})\!\left[1-\dfrac{u_1 R(1-s)}{s(1-u_1)}\right]
&
(1-u_{2,t})\dfrac{(s-u_1)R(1-s)}{s(1-u_1)}
\\[6pt]
u_{2,t}+(1-u_{2,t})\dfrac{u_1 R(1-s)}{s(1-u_1)}
&
(1-s)-(1-u_{2,t})\dfrac{(s-u_1)R(1-s)}{s(1-u_1)}
\end{pmatrix}.
\end{equation}

\noindent For diploids: { \small
\begin{equation}\label{eq:F_dip}
\mathbf F_{\mathrm{dip}}
=\frac{1}{(1-u_1)s}
\begin{pmatrix}
(1-u_{2,t})\!\left[s-u_1R(1-s)\right]
&
(1-u_{2,t})(1-s)R\!\left[s-u_1(1-s)\right]
\\[6pt]
u_1(1-s)R
+u_{2,t}\!\left[s-u_1R(1-s)\right]
&
(1-s)\!\left[(1-R)+u_{2,t}R\right]\!\left[s-u_1(1-s)\right]
+(1-2s)u_1
\end{pmatrix}.
\end{equation}
}

\noindent In both cases, invasion depends on selection $s$, recombination $R$, the resident mutation rate $u_1$, and the time--dependent mutation rate $u_{2,t}$ induced by the invader.  Because $M_2$ has no direct fitness effect, its initial growth is driven entirely by the dynamics of the rare haplotypes $(AM_2,aM_2)$ near the resident mutation--selection balance. Selection at the $A/a$ locus generates the associations that produce the asymptotic multiplicative growth rate of the rare modifier haplotypes, while recombination modulates this rate each generation. The condition $\rho(\mathbf F)>1$ therefore defines the criterion for invasion of the modifier allele $M_2$ near mutation--selection equilibrium.

\subsection{\textbf{Stochastic Mutation Rate ($u_{2, t}$).}} \label{sec:invasion_stochastic_unified}
\noindent We now allow stochasticity to enter through the \emph{transmission process} rather than through selection.  The resident population is assumed to be at the mutation--selection equilibrium \eqref{eq:mut_sel_balance} with forward mutation rate $u_1$, satisfying $0<u_1<s$ in haploids and $s>\tfrac{u_1}{1+u_1}$ in diploids.  These conditions ensure the existence of a mutation--selection balance, so that $a$-alleles persist and indirect selection on the modifier is well defined near this equilibrium.

\noindent Rare modifier allele $M_2$ is introduced whose associated mutation rate is \emph{uniform across the population within each generation} but varies randomly across generations. Stochasticity acts only through time, not across individuals. Specifically, in generation $t$ all $M_2$ carriers experience the same forward mutation rate $u_{2,t}$, with
\[
u_{2,t}\stackrel{\mathrm{i.i.d.}}{\sim}\mathcal L,\qquad 
\mathbb E[u_{2,t}]=u_2,\quad \mathrm{Var}(u_{2,t})=\sigma^2,
\]
where $\mathcal L$ is an arbitrary distribution supported on a biologically admissible interval (e.g.\ $[0,1)$).  No assumption is made about the shape of $\mathcal L$; it may be asymmetric or heavy--tailed, provided moments are finite.

\noindent Linearizing the two--locus dynamics at the resident equilibrium
\[
\hat{\mathbf x}=(\hat x_1,\hat x_2,0,0)^\top
\]
yields the random linear recursion
\[
\bm v_{t+1}=\mathbf F_t(R,u_{2,t})\,\bm v_t,
\qquad 
\bm v_t=(x_{3,t},x_{4,t})^\top,
\]
where $\mathbf F_t$ is given by \eqref{eq:F_hap} or \eqref{eq:F_dip} evaluated at the value of $u_{2,t}$ and fixed recombination rate $R\in[0,\tfrac12]$.  If $\mathbb E[\log\|\mathbf F_t\|]<\infty$, the top Lyapunov exponent
\[
\gamma(R)=\lim_{t\to\infty}\frac{1}{t}\log\big\|\mathbf F_{t-1}\cdots \mathbf F_0\big\|
\]
exists almost surely and is norm--independent \citep{furstenberg1960}. The modifier allele invades if $\gamma(R)>0$, is eliminated if $\gamma(R)<0$, and is marginal at linear order if $\gamma(R)=0$.

\noindent To separate the effects of recombination from those of stochastic mutation rate, we write the random recursion matrix as
\[
\mathbf F_t(R)=\mathbf A_t+R\,\mathbf B_t,
\]
where $\mathbf A_t=\mathbf F_t(0)$ describes the joint action of mutation and selection in the absence of recombination, and $\mathbf B_t$ captures the linear, within–generation effect of recombination.  This decomposition allows us to contrast the cases $R=0$ and $R>0$, and hence to identify how recombination alters invasion when transmission itself is stochastic.

\noindent For haploids, rewriting \eqref{eq:F_hap} gives
\begin{equation}\label{eq:A_B_hap_stoch}
\mathbf A_t^{\mathrm{hap}}=\frac{1}{1-u_1}
\begin{pmatrix}
1-u_{2,t} & 0\\
u_{2,t} & 1-s
\end{pmatrix},
\qquad
\mathbf B_t^{\mathrm{hap}}=\frac{(1-s)(1-u_{2,t})}{s(1-u_1)^2}
\begin{pmatrix}
-u_1 & s-u_1\\
u_1 & -(s-u_1)
\end{pmatrix},
\end{equation}
and for diploids,
\begin{equation}\label{eq:A_B_dip_stoch}
\mathbf A_t^{\mathrm{dip}}=\frac{1}{(1-u_1)s}
\begin{pmatrix}
(1-u_{2,t})\,s & 0\\
u_{2,t}\,s & (1-s)\,[s-u_1(1-s)]+(1-2s)u_1
\end{pmatrix},
\end{equation}
\[
\mathbf B_t^{\mathrm{dip}}=\frac{(1-s)(1-u_{2,t})}{s(1-u_1)}
\begin{pmatrix}
-u_1 & s-u_1(1-s)\\
u_1 & -[s-u_1(1-s)]
\end{pmatrix}.
\]

\noindent In both haploids and diploids, $\mathbf A_t$ coincides with the no–recombination system, while $\mathbf B_t$ has zero column sums, reflecting the fact that recombination reshuffles haplotypes among genetic backgrounds without changing their total frequency.  

\noindent
This structure highlights the special role of $R=0$. In the absence of recombination, the dynamics reduce to products of lower--triangular random matrices, so the top Lyapunov exponent $\gamma(0)$ admits a closed--form expression given by the time average of the logarithm of the dominant diagonal entry. In this case, invasion is governed by a one--dimensional multiplicative process and depends only on the marginal distribution of mutation rates.

\noindent For any $R>0$ with $\mathbb P(u_{2,t}>0)=1$, by contrast, recombination couples modifier alleles and background states so that all entries of $\mathbf F_t(R)$ are strictly positive infinitely often. The resulting sequence of matrices is therefore \emph{primitive}, guaranteeing the existence and uniqueness of the top Lyapunov exponent \citep{furstenberg1960}. Although recombination enters linearly within each generation through $R\,\mathbf B_t$, its effect on invasion is mediated by products of non--commuting random matrices across generations. As a result, the dependence of $\gamma(R)$ on $R$ is generally nonlinear and need not be monotone.

\noindent
This distinction between $R=0$ and $R>0$ is fundamental. In the absence of recombination, stochasticity in transmission influences invasion solely through the geometric mean of marginal growth factors. When $R>0$, by contrast, temporal variation interacts with genetic mixing and other evolutionary parameters to modify the structure of long--run growth itself, so that the distribution of $u_{2,t}$, recombination, selection, and the resident mutation rate may affect not only the rate but also the direction of selection on modifiers.

\paragraph{\textbf{Baseline at $R=0$.}}  We first analyze the case of complete linkage. When $R=0$, the random recursion matrices $\mathbf F_t(0)$ are lower triangular, and their products remain lower triangular. The top Lyapunov exponent is therefore given by the larger of the time–averaged logarithms of the diagonal entries.

\begin{itemize}
\item \textit{Haploids.}  At $R=0$,
\[
\mathbf F_t(0)=\frac{1}{1-u_1}
\begin{pmatrix}
1-u_{2,t} & 0\\
u_{2,t} & 1-s
\end{pmatrix},
\]
and hence
\[
\gamma(0)
=
-\log(1-u_1)
+
\max\!\Big\{
\mathbb E[\log(1-u_{2,t})],\;
\log(1-s)
\Big\}.
\]

\item \textit{Diploids.}  At $R=0$,
\[
\mathbf F_t(0)=\frac{1}{1-u_1}
\begin{pmatrix}
1-u_{2,t} & 0\\
u_{2,t} & 1-s(1+u_1)
\end{pmatrix},
\]
and, provided $1-s(1+u_1)>0$,
\[
\gamma(0)
=
-\log(1-u_1)
+
\max\!\Big\{
\mathbb E[\log(1-u_{2,t})],\;
\log\!\big(1-s(1+u_1)\big)
\Big\}.
\]
\end{itemize}
\noindent The diagonal entries correspond to the instantaneous eigenvalues
\[
\lambda_1(t)=\frac{1-u_{2,t}}{1-u_1},
\qquad
\lambda_2=
\begin{cases}
\dfrac{1-s}{1-u_1}, & \text{haploids},\\[6pt]
\dfrac{1-s(1+u_1)}{1-u_1}, & \text{diploids}.
\end{cases}
\]
\noindent Under admissible mutation--selection balance conditions, $\lambda_2<1$ in both haploids and diploids. Hence $\lambda_2$ cannot generate positive growth in the frequency of $M_2$. Invasion must therefore depend on the transmission eigenvalue $\lambda_1(t)$. The invasion condition is
\begin{equation}
\label{eq:stochastic_R0_inv_condition}
\gamma(0)>0
\;\Longleftrightarrow\;
\mathbb E[\log\lambda_1(t)]>0
\;\Longleftrightarrow\;
\mathbb E[\log(1-u_{2,t})]>\log(1-u_1).
\end{equation}
\noindent Thus, the direction of selection on the rare modifier allele is determined entirely by the sign of $\mathbb E[\log(1-u_{2,t})]-\log(1-u_1)$. If this quantity is positive, $M_2$ increases exponentially when rare; if it is negative, $M_2$ is eliminated exponentially. Equality corresponds to neutral stability to first order. 

\noindent Because $\log(1-x)$ is strictly concave on $[0,1)$, Jensen’s inequality implies
\[
\mathbb E[\log(1-u_{2,t})]\le \log\!\big(1-\mathbb E[u_{2,t}]\big),
\]
with equality only when $u_{2,t}$ is constant (i.e.\ $\sigma^2=0$). Temporal variance in the mutation rate therefore reduces the long--run multiplicative transmission of allele $A$ relative to a deterministic modifier allele with the same mean. Consequently, even when $\mathbb E[u_{2,t}]<u_1$, sufficiently large variance can prevent invasion.

\noindent When condition~\eqref{eq:stochastic_R0_inv_condition} holds, the Lyapunov exponent simplifies to
\[
\gamma(0)=\mathbb E[\log(1-u_{2,t})]-\log(1-u_1),
\]
which is independent of the selection coefficient $s$ and of ploidy. Appendix~\ref{appendix:approximation} and Table~\ref{tab:lyap_G_summary_landscape} summarize $\mathbb E[\log(1-u_{2,t})]$ for several distributions and relate it to $\log(1-u_1)$, and Appendix \ref{appendix:lyap_acc}, analyzes accuracy of these estimations.

\noindent If instead
\[
\mathbb E[\log(1-u_{2,t})]\le \log(1-u_1),
\]
then $\mathbb E[\log\lambda_1(t)]\le 0$. Since $\lambda_2<1$ under mutation--selection balance, both diagonal growth rates are non--positive, and
\[
\gamma(0)
=
\max\!\big\{\mathbb E[\log\lambda_1(t)],\,\log\lambda_2\big\}
<0.
\]

\noindent In this case invasion is not possible. The eigenvalue $\lambda_2$ does not provide an alternative invasion criterion; it only determines the exponential rate at which the modifier allele is eliminated when the transmission term fails to generate growth. In this regime, $\gamma(0)$ is independent of the distribution of $u_{2,t}$ and depends only on selection, ploidy, and the resident mutation rate. Stronger selection decreases $\lambda_2$ and accelerates loss of the modifier allele, while increasing $u_1$ weakens this constraint in haploids and has a sign--dependent effect in diploids.

\noindent In summary, with complete linkage ($R=0$), invasion is possible only through differences in long--run multiplicative transmission. Temporal variability in mutation rate reduces this effect by Jensen’s inequality, shrinking the invasion domain relative to the deterministic case. When this condition fails, selection enforces extinction, and stochasticity in mutation rates plays no further role.

\subparagraph{\textbf{Role of selection and resident mutation rate at $R=0$.}} The influence of the remaining parameters on $\gamma(0)$ follows directly from the explicit expressions derived above.
\noindent When
\[
\mathbb E[\log(1-u_{2,t})]>\log(1-u_1),
\]
the Lyapunov exponent is
\[
\gamma(0)=\mathbb E[\log(1-u_{2,t})]-\log(1-u_1),
\]
and does not depend on the selection coefficient $s$. In this case, increasing the resident mutation rate $u_1$ strictly increases $\gamma(0)$, because $-\log(1-u_1)$ is increasing in $u_1$. Thus, conditional on invasion, larger values of $u_1$ increase the exponential growth rate of $M_2$; conversely, when $\gamma(0)<0$, they increase its exponential rate of decline.

\noindent When
\[
\mathbb E[\log(1-u_{2,t})]\le \log(1-u_1),
\]
the Lyapunov exponent is instead
\[
\gamma(0)=\log\lambda_2<0,
\]
and its value is independent of the distribution of $u_{2,t}$. In this regime, $\gamma(0)$ is determined entirely by selection, ploidy, and the resident mutation rate. For haploids,
\[
\lambda_2=\frac{1-s}{1-u_1},
\]
so $\gamma(0)$ decreases monotonically with $s$ and increases monotonically with $u_1$. For diploids,
\[
\lambda_2=\frac{1-s(1+u_1)}{1-u_1},
\]
so $\gamma(0)$ decreases with $s$, while its dependence on $u_1$ changes sign at $s=\tfrac12$, subject to the admissibility condition $1-s(1+u_1)>0$.

\noindent  Thus, with complete linkage, the comparison between $\mathbb E[\log(1-u_{2,t})]$ and $\log(1-u_1)$ determines the sign of $\gamma(0)$ and hence the direction of change of $M_2$'s frequency when $M_2$ is rare. The parameters $s$ and $u_1$ affect only the magnitude of $\gamma(0)$, scaling the rate of increase or decrease without altering the invasion criterion.

\noindent \subparagraph{Simulation Results.} We use numerical iteration to examine further how temporal variability in mutation rate affects the evolutionary fate of a modifier allele when recombination is absent ($R=0$). Within each generation, all carriers of the modifier $M_2$ experience the same mutation rate, but this rate varies independently across generations. Formally, $M_2$ induces an i.i.d.\ sequence $\{u_{2,t}\}_{t\ge0}$ with
\[
u_{2,t}\sim\mathcal L(u_2,\sigma^2), \qquad
\mathbb E[u_{2,t}]=u_2,\quad \mathrm{Var}(u_{2,t})=\sigma^2,
\]
supported on $[0,1)$. No assumption beyond bounded support is imposed, which ensures that mutation probabilities remain  biologically meaningful.

\noindent To isolate the effect of temporal variability from that of the mean, we compare four distributional families for $\mathcal L$: uniform, beta, truncated log-normal, and truncated gamma. Each family is parameterized to match a common mean $u_2$ and variance $\sigma^2$ (Appendix~\ref{appendix:approximation}). The distributions differ only in how probability mass is allocated over $[0,1]$, thereby representing distinct modes of temporal fluctuation while holding the first two moments fixed.

\noindent Under complete linkage $(R=0)$, invasion is governed by the sign of
\[
\gamma(0)=\mathbb E[\log(1-u_{2,t})]-\log(1-u_1),
\]
so that $M_2$ increases when $\mathbb E[\log(1-u_{2,t})]>\log(1-u_1)$. When $u_{2,t}$ is constant ($u_{2,t}\equiv u_2$) (see Appendix  \ref{sec:invasion_deterministic_unified}), this reduces to the classical condition $u_2<u_1$. When $u_{2,t}$ fluctuates, Jensen’s inequality implies
\[
\mathbb E[\log(1-u_{2,t})]\le \log(1-u_2),
\]
with equality only in the absence of variability. Thus, with a fixed mean $u_2$, temporal variability always lowers $\gamma(0)$. A modifier allele that reduces the mean mutation rate may therefore fail to invade if its mutation rate fluctuates sufficiently, while an allele that increases the mean mutation rate cannot invade and is eliminated more rapidly as variability increases.

\noindent With complete linkage, the roles of the remaining parameters are transparent.  The invasion condition depends on the resident mutation rate $u_1$ only through the reference value $\log(1-u_1)$ and is independent of the selection coefficient $s$ once a mutation--selection equilibrium exists.  Increasing $u_1$ lowers $\log(1-u_1)$, thereby relaxing the invasion condition and expanding the set of $(u_2,\sigma^2)$ for which $M_2$ can increase, whereas smaller $u_1$ makes invasion uniformly more restrictive.  Changes in $s$ affect mutation-selection equilibrium allele frequencies and the speed of convergence, but do not alter the sign of $\gamma(0)$ at $R=0$. Accordingly, the simulations fix $s$ and $u_1$ and vary $(u_2,\sigma^2)$ to isolate the evolutionary consequences of temporal variability and distributional shape.

\begin{figure}[t]
    \centering
    \includegraphics[width=\textwidth]{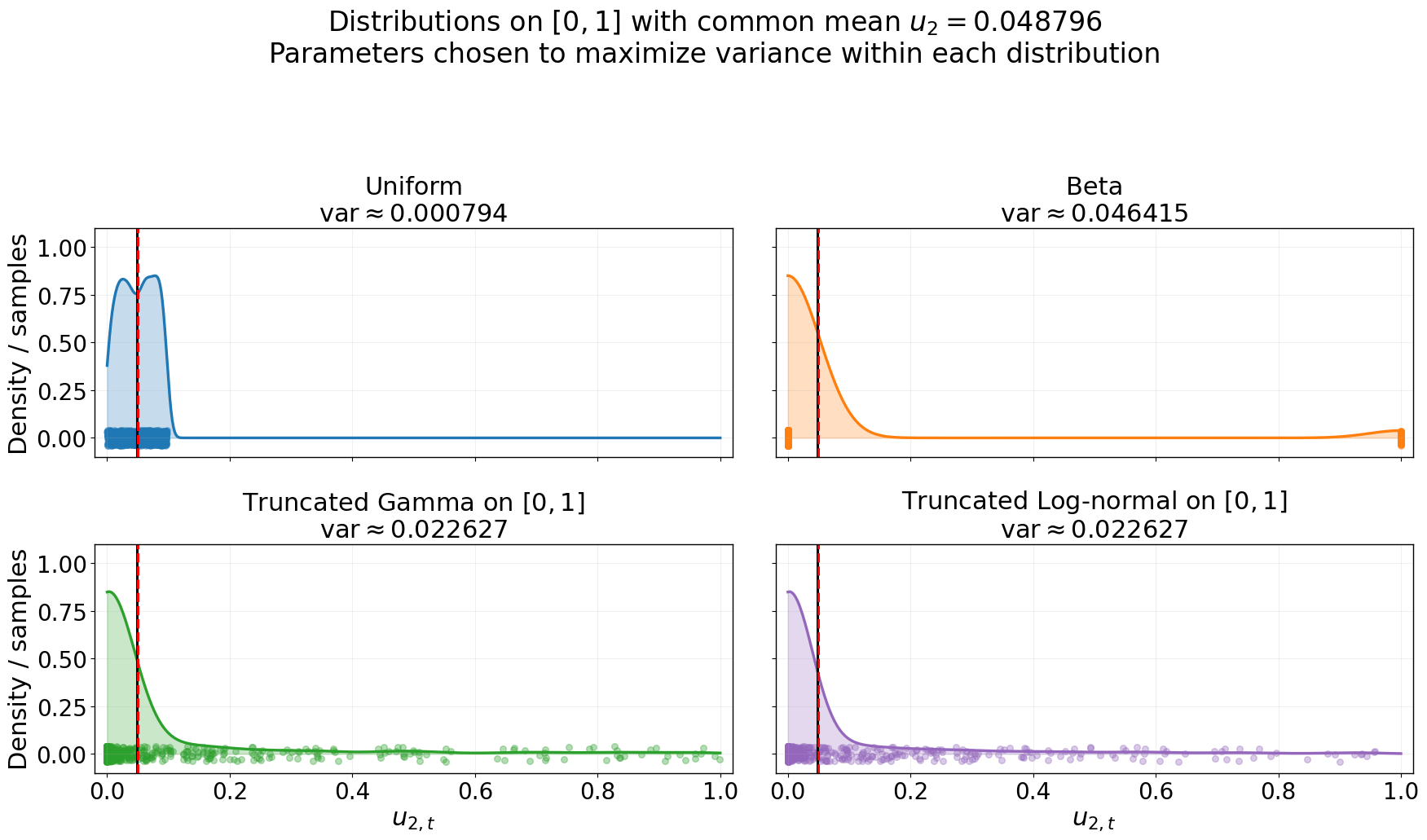}
    \caption{
    \textbf{ Distributions on $[0,1]$ with a common mean $u_2 = 0.048796$. } Points show simulated realizations of $u_{2,t}$ and solid curves show kernel density estimates. Panels correspond to uniform, beta, truncated gamma, and truncated log–normal distributions. In each case, parameters are chosen to produce the largest variance attainable within the given family under the constraint of support on $[0,1]$ and fixed mean (up to numerical precision). Reported variance values illustrate the wide disparity in admissible temporal variability across distributions despite identical means and bounds.}
    \label{fig:max_variance_families}
\end{figure}
\noindent Figure~\ref{fig:max_variance_families} illustrates how strongly the feasible range of temporal variability depends on distributional form. The uniform distribution admits only limited variance once the mean and support are fixed. The beta distribution can achieve much larger variance by concentrating mass near the boundaries, particularly near $u_{2,t}=1$. The truncated log-normal and truncated gamma distributions are right-skewed and occupy an intermediate position: both place most mass near small $u_{2,t}$, but differ in how much weight they assign to moderate and large values after truncation. As summarized in Table~\ref{tab:distribution-summary}, these differences impose distribution-specific bounds on the combinations of $(u_2,\sigma^2)$ that are feasible. 

\begin{figure}[t]
    \centering
    \includegraphics[width=\textwidth]{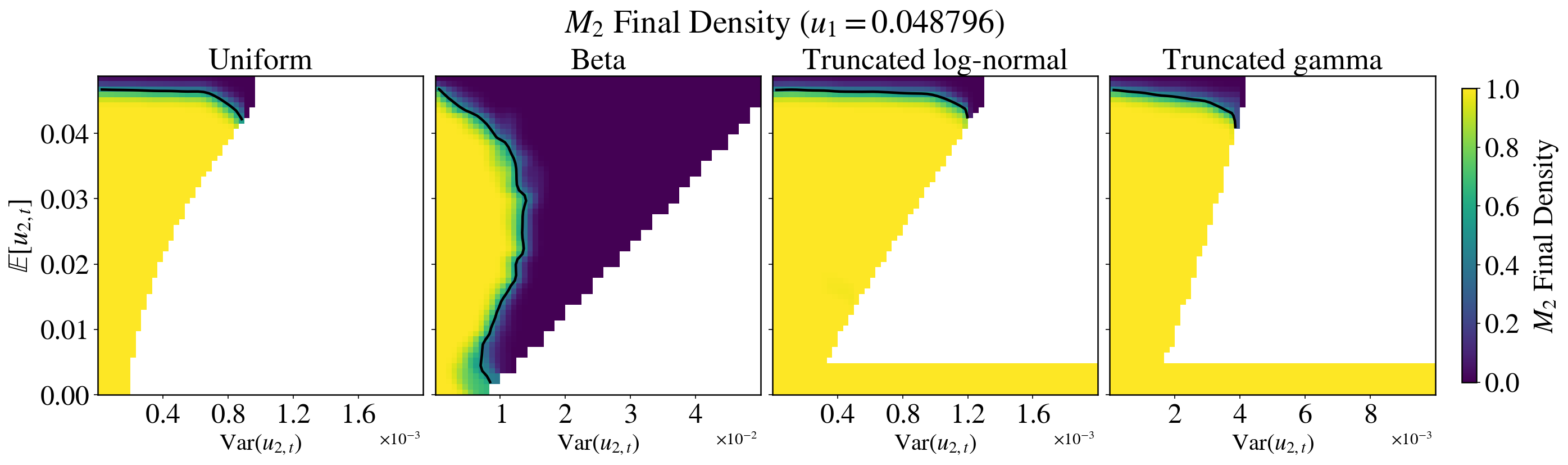}
    \caption{
   \textbf{ Fixation probability of a modifier allele under temporally fluctuating mutation rates.} Panels correspond to the distribution governing the across-generation mutation rate $u_{2,t}$ induced by $M_2$ (uniform, beta, truncated log-normal, truncated gamma). Axes show $\mathrm{Var}(u_{2,t})$ and $\mathbb{E}[u_{2,t}]$. Color denotes the final density of the modifier allele $M_2$ at generation $5000$, averaged across replicate simulations. The black contour marks the $50\%$ fixation boundary.  All simulations assume complete linkage ($R=0$), selection coefficient $s=0.2$, nd baseline mutation rate $u_1=0.048796$.}
    \label{fig:fixation_probability_fluctuations}
\end{figure}

\noindent In each simulation, the population is initialized at the deterministic mutation--selection equilibrium of the resident modifier $M_1$ in the absence of $M_2$. The resident mutation rate is fixed at $u_1=0.048796$. This value should be interpreted as an effective mutation rate at the scale of the genetic unit influenced by the modifier—such as a block of tightly linked sites or a genome-wide mutation process—rather than as a single-site mutation probability. At this aggregate scale, mutation rates of order $10^{-2}$ are appropriate and yield interior equilibria under mutation--selection balance. The selection coefficient is set to $s=0.2$ to ensure that selection on modifiers is detectable over finite time. Recombination is set to zero ($R=0$) to isolate the effects of temporal variability under complete linkage, the regime in which the analytical criterion is exact.

\noindent The modifier $M_2$ is introduced at frequency $10^{-4}$, equally distributed across genetic backgrounds, approximating the rare-allele limit assumed by the invasion analysis while avoiding numerical degeneracy. The full two-locus recursion is iterated for $5{,}000$ generations. This time horizon is sufficient for trajectories subject to sustained directional selection to exhibit clear monotonic changes in frequency. Extending simulations beyond this point does not alter the qualitative ordering of outcomes, but only increases separation along the same trajectories. Accordingly, colors represent the final density of $M_2$ haplotypes at generation $5{,}000$ and provide a finite-time summary of the direction and strength of selection on the modifier allele. White regions correspond to $(u_2,\sigma^2)$ pairs that are infeasible for the given distribution under truncation and moment matching. 

\noindent Simulation outcomes are summarized in Fig.~\ref{fig:fixation_probability_fluctuations}. Across all distributions, outcomes are organized by the geometric-mean criterion. When $u_2<u_1$, distributions that place little probability mass at large mutation rates—such as the uniform and truncated log-normal—maintain relatively large values of $\mathbb E[\log(1-u_{2,t})]$ and permit substantial increases in the frequency of $M_2$ haplotypes. In contrast, distributions that realize variance by allocating mass to large $u_{2,t}$—most prominently the beta distribution and high-variance truncated gamma—strongly depress the logarithmic mean and yield low final frequencies even when the arithmetic mean mutation rate is reduced. When $u_2>u_1$, $M_2$ declines under all distributions, and increasing variability accelerates this decline.

\noindent The mechanism underlying these differences is captured by the upper-tail probability
\[
\xi(u_1)=\Pr(u_{2,t}>u_1).
\]
\noindent Because $\log(1-u)$ is strictly concave with slope diverging as $u\uparrow1$, realizations near the upper boundary contribute disproportionately to $\mathbb E[\log(1-u_{2,t})]$. For fixed $(u_2,\sigma^2)$, distributions that generate variance by increasing $\xi(u_1)$ produce smaller logarithmic means and hence smaller $\gamma(0)$. As emphasized in Table~\ref{tab:distribution-summary}, invasion therefore depends not only on the magnitude of temporal variance but on how that variance is distributed relative to $u_1$.

\noindent
This dependence is quantified in Appendix~\ref{appendix: compare_derivations}, Table~\ref{tab:u2star_comparison}, which reports the critical mean $u_2^\ast$ satisfying $\mathbb E[\log(1-u_{2,t})]=\log(1-u_1)$ at fixed $\sigma^2$. Because the variance is held constant, differences in $u_2^\ast$ arise solely from the shape of the distribution. Distributions that allocate more probability to large mutation rates admit invasion only for smaller $u_2$, while those concentrating variability at small rates permit invasion at higher $u_2$.

\noindent
Taken together, the simulations confirm the analytical prediction for $R=0$: selection on mutation-rate modifiers is determined by the logarithmic mean of $(1-u_{2,t})$, not by the arithmetic mean of $u_{2,t}$. Temporal variability systematically reduces this logarithmic mean, and differences in how variability is realized can qualitatively alter evolutionary outcomes even when mean effects are identical.
\begin{table}[t]
\centering
\renewcommand{\arraystretch}{1.18}
\setlength{\tabcolsep}{6pt}

\caption{\textbf{How distributional shape translates temporal variance into selection at $R=0$.}
At fixed $(u_2,\sigma^2)$, invasion depends on $\mathbb E[\log(1-u_{2,t})]$, which is most sensitive to probability mass for large mutation rates.
For each family we summarize: (i) how variance is realized at fixed mean, (ii) the resulting behavior of $\xi(u_1)=\Pr(u_{2,t}>u_1)$, and (iii) the corresponding patterns in Figs.~\ref{fig:max_variance_families} and \ref{fig:fixation_probability_fluctuations}.}

\begin{tabular}{
>{\raggedright\arraybackslash}p{0.17\textwidth}
>{\raggedright\arraybackslash}p{0.39\textwidth}
>{\raggedright\arraybackslash}p{0.36\textwidth}
}
\toprule
\textbf{Distribution} &
\textbf{How variance is generated at fixed mean $u_2$} &
\textbf{Implications for $\mathbb E[\log(1-u_{2,t})]$ and simulation patterns} \\
\midrule

Uniform &
Variance is determined by interval width; increasing $\sigma^2$ widens support around $u_2$ subject to $[0,1]$. Upper-tail mass is limited by geometry. &
$\xi(u_1)$ increases slowly with $\sigma^2$, so $\mathbb E[\log(1-u_{2,t})]$ declines gradually. The invasion boundary in Fig.~\ref{fig:fixation_probability_fluctuations} is weakly dependent on $\sigma^2$.\\[5pt]

Beta &
Large $\sigma^2$ is realized by shifting probability toward both boundaries, including substantial mass near $u_{2,t}\approx 1$. &
$\xi(u_1)$ rises rapidly with $\sigma^2$, leading to a pronounced reduction in
$\mathbb{E}\!\left[\log(1 - u_{2,t})\right]$. Even when $u_2 < u_1$, high-$M_2$
realizations sharply compress the
$\mathbb{E}\!\left[\log(1 - u_{2,t})\right]$ term as variance increases.\\[5pt]

Truncated log-normal &
Strong right skew. For small means, large $\sigma^2$ is attainable while most mass remains near zero; truncation limits (but does not eliminate) the upper tail. &
Relative to beta, $\xi(u_1)$ grows more slowly, so $\mathbb E[\log(1-u_{2,t})]$ remains larger at comparable $(u_2,\sigma^2)$. The extended yellow region at low $u_2$ reflects that large variances are feasible at small means in this family, not a dynamical threshold.\\[5pt]

Truncated gamma &
Right skew with more weight at intermediate-to-large values than the truncated log-normal under comparable moment constraints. Large $\sigma^2$ is also attainable at low $u_2$. &
For the same $(u_2,\sigma^2)$, $\xi(u_1)$ is larger than under the truncated log-normal, yielding smaller $\mathbb E[\log(1-u_{2,t})]$ and a more restrictive invasion region. The extended yellow region at low $u_2$ again reflects feasibility (large attainable $\sigma^2$ at small means).\\

\bottomrule
\end{tabular}
\label{tab:distribution-summary}
\end{table}
\noindent \paragraph{\textbf{Stochastic mutation rate with recombination.} }For $R>0$ and admissible values of $s$ and $u_1$, the per-generation invasion matrix $\mathbf F_t(R)$ is entrywise nonnegative, with strictly positive off-diagonal entries whenever $u_{2,t}>0$. As a consequence, recombination couples the rare haplotypes each generation, and the products $\prod_{t=0}^{T-1}\mathbf F_t(R)$ generally do not commute across generations. The long-run growth rate of the modifier is therefore governed by a Lyapunov exponent,
\[
\gamma(R)\;=\;\lim_{T\to\infty}\frac1T\log\bigl\|\mathbf F_{T-1}(R)\cdots \mathbf F_0(R)\bigr\|,
\]
which has no closed form in general. Recombination has no direct fitness effect at the modifier locus; it acts indirectly by reshaping, from generation to generation, the stochastic associations between $M_2$ and the selected backgrounds generated by mutation and selection. This contrasts sharply with complete linkage. 
\noindent To investigate how $u_1$, $\mathcal L(u_2, \sigma^2)$, $s$, and $R$ interact for $R>0$, we analyze the system numerically. Suppose $M_2$–associated mutation rate follows
\[
u_{2,t}\stackrel{\text{i.i.d.}}{\sim}\mathrm{Beta}(\alpha,\beta),
\]
with mean $u_2=0.04$ and variance
\[
\sigma^2 \in \{0,\,0.25,\,0.50,\,0.75,\,0.95\}\times u_2(1-u_2),
\qquad u_2(1-u_2)=0.0384.
\]
\noindent Scaling $\sigma^2$ by $u_2(1-u_2)$—the maximal variance for any random variable on $[0,1]$ with mean $u_2$—allows comparisons across variance levels on a common, interpretable scale.

\noindent The value $\mathbb{E}[u_{2,t}] = u_2=0.04$ (and $u_1=0.05$) should be interpreted as an \emph{effective mutation rate} at the scale influenced by the modifier (e.g.\ a tightly linked block or genome-wide process), not as a single-site rate. At this aggregate scale, mutation rates of order $10^{-2}$ are biologically plausible and permit a mutation–selection balance.

\noindent The resident mutation rate is fixed at $u_1=0.05$, and we consider two selection regimes:
\[
s\in\{0.06,\,0.20\},
\]
representing weak and strong selection on the selected background, respectively.

\noindent For each variance level, we draw a single realization of $\{u_{2,t}\}$ and reuse it across recombination values and ploidies to reduce Monte Carlo noise and isolate structural effects. For $R$ on a fine grid in $[0,\tfrac12]$, we iterate
\[
\bm v_{t+1}=\mathbf F_t(R)\,\bm v_t,
\]
renormalizing by the $\ell^1$ norm at each step. By the Furstenberg–Kesten theorem, the time–average of the logarithmic rescaling converges almost surely to the top Lyapunov exponent $\gamma(R)$ \cite{furstenberg1960}. 

\begin{figure}[t]
    \centering
    \includegraphics[width=\textwidth]{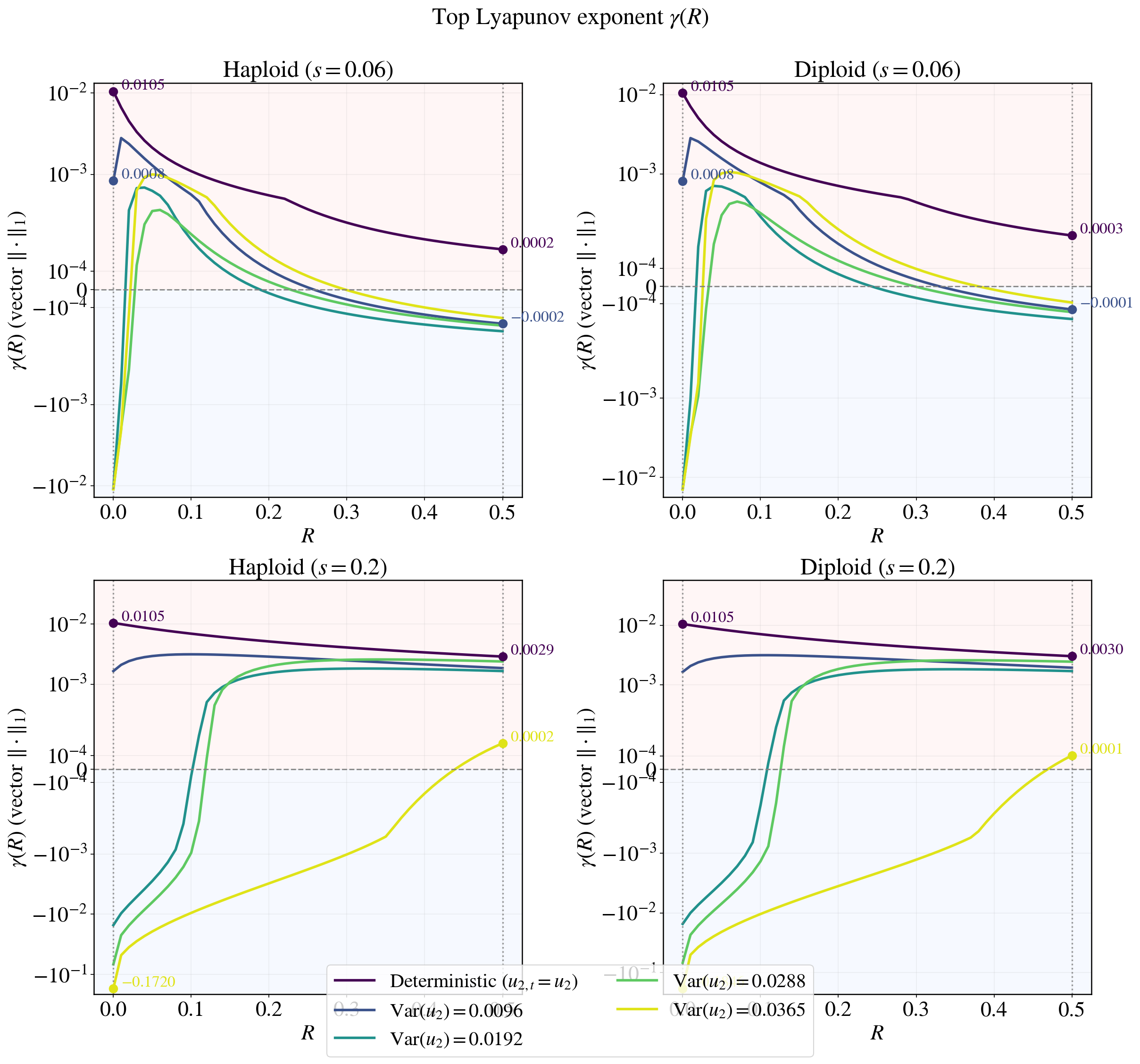}
    \caption{
    \textbf{Top Lyapunov growth rate $\gamma(R)$ as a function of recombination rate $R$}.
    For haploid (left column) and diploid (right column) populations under
    stochastic mutation. Mutation rates $u_{2,t}$ are i.i.d.\ Beta-distributed with fixed mean $\mathbb{E}[u_{2,t}] = 0.04$ and varying variance $\operatorname{Var}(u_2)$; the deterministic case corresponds to $\operatorname{Var}=0$.  Curves show the top Lyapunov exponent of the random matrix product. Blue and red shaded regions indicate $\gamma(R)<0$ and $\gamma(R)>0$, respectively. Vertical dotted lines mark $R=0$ and $R=1/2$. Selection coefficients are $s=0.06$ (top row) and $s=0.20$ (bottom row).
    }
    \label{fig:gamma_recombination}
\end{figure} 

\noindent Figure~\ref{fig:gamma_recombination} summarizes the resulting qualitative
structure. With complete linkage, the sign of $\gamma(0)$ is invariant across selection regimes and ploidies, and changes only when the variance of the mutation-rate distribution is altered. No change in $s$, ploidy, or $u_1$ can reverse this sign at $R=0$; only the distribution of $u_{2,t}$ enters the invasion criterion \eqref{eq:stochastic_R0_inv_condition}.

\noindent When $R>0$, the situation is more complicated. Because growth is multiplicative across generations, recombination converts the problem into one of noncommuting random matrix products, and $\gamma(R)$ becomes sensitive to parameters that are neutral with complete linkage. Recombination can therefore change not only the magnitude but also the \emph{direction} of selection on the modifier.

\noindent
This qualitative shift is illustrated in Fig.~\ref{fig:gamma_recombination}. For $R>0$, even modest increases in recombination are sufficient to permit changes in invasion outcome driven by selection regime or ploidy, a dependence that does not occur for $R=0$. As a result, invasion is no longer determined solely by the mutation process, but by the joint effects of recombination, background selection, and temporal mutation variability.

\noindent Thus, temporal variability in $u_{2,t}$ creates a regime in which invasion criteria depend jointly on recombination, selection strength, and resident mutation rate. Such sign reversals are excluded in deterministic models and under complete linkage, but arise generically once stochastic mutation and recombination are combined.

\noindent
\subparagraph{\textbf{Invasion dependence on Parameters.}} Invasion of $M_2$ is determined by the sign of $\gamma(R)$. At complete linkage ($R=0$), $\gamma(0)$ reduces to the scalar log-mean condition \eqref{eq:lyap_clean}, so invasion depends only on $u_1$ and the distribution of $u_{2,t}$ through $\mathbb E[\log(1-u_{2,t})]$. For $R>0$, $\gamma(R)$ is the top Lyapunov exponent of a noncommuting random matrix product; hence invasion generally depends on all parameters that shape the modifier’s genetic background (including $s$ and ploidy), and neither monotonicity nor a closed form should be assumed.
\begin{table}[t]
\centering
\caption{\textbf{How parameters enter the invasion condition $\gamma(R)\gtrless 0$.} At $R=0$, the sign is fixed by \eqref{eq:stochastic_R0_inv_condition}; for $R>0$, parameters act through modifier--background associations in the random matrix product and can change both the sign and the shape of $\gamma(R)$.}
\label{tab:param_dependence}
\begin{tabular}{p{2.5cm} p{5.3cm} p{6.0cm}}
\hline
\textbf{Parameter} & \textbf{Effect on invasion at $R=0$} & \textbf{Effect on invasion when $R>0$} \\
\hline
$u_1$ &
Enters additively as $-\log(1-u_1)$; increasing
$u_1$ increases $\gamma(0)$ and can reverse invasion. &
Affects $\mathbf F_t(R)$ directly and alters the resident background on which
$M_2$ recombines; varying $u_1$ can change the sign of $\gamma(R)$ and shift any
critical threshold (e.g.\ $R^\ast$). \\[0.35em]

Distribution of $u_{2,t}$&
Enters only through $\mathbb E[\log(1-u_{2,t})]$; invasion occurs if
$\mathbb E[\log(1-u_{2,t})]>\log(1-u_1)$.&
Enters through the full temporal sequence in the matrix product; features beyond
the log-mean (e.g.\ tail mass near $1$) can change the sign of $\gamma(R)$ and
induce non-monotone dependence on $R$. \\[0.35em]

$u_2=\mathbb E[u_{2,t}]$ &
Does not determine invasion by itself except in the deterministic limit
($u_{2,t}\equiv u_2$), where invasion is $u_2<u_1$. &
Shifts the central tendency of $u_{2,t}$ and can change the sign of $\gamma(R)$,
but does not fix it without additional distributional information. \\[0.35em]

$\sigma^2=\mathrm{Var}(u_{2,t})$ &
Affects invasion only insofar as it changes $\mathbb E[\log(1-u_{2,t})]$
(Jensen); variance alone is not the criterion. &
Can shift or create invasion reversals by changing the frequency and severity of
high-$u_{2,t}$ episodes; variance alone does not predict the qualitative response
of $\gamma(R)$ without information on the distribution’s shape. \\[0.35em]

$s$ &
Does not enter \eqref{eq:stochastic_R0_inv_condition}; cannot change the sign of
$\gamma(0)$.&
Alters how strongly selection amplifies or purges backgrounds with which $M_2$
associates; varying $s$ can change the sign of $\gamma(R)$ and shift thresholds
(e.g.\ $R^\ast$), and may change the monotonicity of $\gamma(R)$ in $R$. \\[0.35em]

Ploidy &
Does not enter \eqref{eq:stochastic_R0_inv_condition}; cannot change the sign of
$\gamma(0)$.&
Changes the genotype-to-fitness mapping and hence the conditional background
composition of $M_2$; for fixed $(u_1,\text{law}(u_{2,t}),s)$, ploidy can change
the sign of $\gamma(R)$ and the location of thresholds. \\[0.35em]

$R$ &
At $R=0$ there is no variation in recombination by definition; $R$ has no
separate effect on $\gamma(0)$. &
Controls the persistence time of modifier--background associations; varying $R$
can change the sign of $\gamma(R)$ and need not do so monotonically. \\
\hline
\end{tabular}
\end{table}

\noindent
Table~\ref{tab:param_dependence} isolates the distinction between complete linkage, where invasion is fixed by a one-dimensional log-mean comparison, and incomplete linkage, where invasion depends on the full stochastic dynamics of modifier--background associations and can exhibit threshold and non-monotone responses to parameters.

\noindent
Since the dependence of invasion on parameters that shape modifier–background associations (including $R$, $s$, $u_1$, and ploidy) can be nonlinear and non-monotone, and invasion boundaries must in general be defined implicitly. We formalize this notion by defining parameter-specific critical thresholds (e.g.\ a recombination cutoff $R^\ast$) and provide a systematic numerical procedure for their computation. Detailed definitions, numerical methods, and worked examples illustrating non-monotone dependence and threshold behavior are given in Appendix~\ref{appendix:parameter_dependence}.

\newpage
\section*{Discussion}
Most analyses of modifier evolution in large populations introduce stochasticity through fluctuating selection while holding transmission parameters fixed. Here we reverse this emphasis. Genotypic viabilities are constant, but transmission is allowed to vary: a selectively neutral modifier allele alters the mutation rate at a linked selected locus, either deterministically or stochastically across generations. Our aim is not to provide an exhaustive catalog of outcomes, but to identify which features of the system determine whether a modifier allele invades, and which merely rescale the rate at which invasion proceeds.

\noindent When the modifier allele induces a constant mutation rate, the invasion problem is structurally simple. In both haploid and diploid models, invasion depends only on the ordering of mutation rates: a modifier allele increases when rare if it lowers the mutation rate relative to the resident and is lost if it increases that rate. This is the classical Reduction Principle. Recombination does not alter this conclusion. Its effect is purely quantitative, weakening indirect selection on the modifier by moving the leading eigenvalue of the linearized system toward unity, but never changing its sign. Selection coefficients, dominance, ploidy, and recombination therefore influence how fast invasion occurs, not whether it occurs.

\noindent Allowing the modifier-induced mutation rate to fluctuate across generations alters the invasion criterion even under complete linkage. When $R=0$, invasion is governed by a geometric–mean condition: what matters is the expected logarithmic growth rate of the invading modifier allele across generations, not its arithmetic mean mutation rate. Temporal variance in transmission necessarily reduces this geometric mean relative to a constant rate with the same average, because rare generations with very high mutation have a disproportionate negative effect on long-term growth. Consequently, a modifier allele that lowers the \emph{mean} mutation rate may nonetheless fail to invade if that reduction is achieved through sufficiently variable transmission.

\noindent A modifier allele that increases the mean mutation rate still cannot invade. Rather, stochastic transmission extends the Reduction Principle by weakening its sufficiency: mean reduction alone is no longer enough. Selection favors modifier alleles that reduce mutation rates reliably across generations, not merely on average. Temporal variance thus shrinks the region of parameter space in which invasion is possible by lowering $\mathbb{E}[\log(1-u_{2,t})]$ relative to a deterministic rate with the same mean.

\noindent In the presence of recombination, stochasticity in transmission has qualitatively new consequences. With $R>0$, recombination reshuffles the genetic associations produced each generation by selection and mutation, thereby changing how temporal variation in transmission is translated into long-term growth. In contrast to the deterministic case, recombination can now either increase or decrease the invasion exponent, depending on over which generations it effectively averages. Breaking associations formed during high–mutation generations can raise the geometric mean growth rate, while breaking those formed during low–mutation generations can lower it. As a result, the dependence of invasion on recombination need not be monotonic, and modifier alleles with identical mean mutation rates can differ in invasibility solely because they differ in the variance or tail behavior of their mutation-rate distributions.

\noindent Importantly, recombination is not the only parameter that can have qualitative effects in this regime. When transmission is stochastic and $R>0$, invasion depends jointly on mutation rates, selection strength, ploidy, and the full distribution of $u_{2,t}$. Parameters that were merely quantitative factors under deterministic transmission can now change the sign of selection on the modifier. Mean mutation rates alone are therefore not the sole predictors of evolutionary outcome once temporal variability and recombination are both present.

\noindent These results suggest a possible, though speculative, perspective on why mutation rates are not driven to zero. Classical explanations invoke a drift barrier, costs of replication fidelity, or physiological constraints. Our results point to an additional mechanism: if reduction in the mean mutation rate is accompanied by increased temporal variance, then selection acts on the geometric mean fidelity rather than the arithmetic mean. Thus, the lowest evolvable mutation rate would be limited not by its mean value alone, but by the variability required to achieve it. We emphasize, however, that this interpretation goes beyond the present model.

\noindent Several extensions could be valuable. Introducing temporal autocorrelation in $u_{2,t}$ would allow high–mutation generations to cluster, potentially amplifying or dampening their effect on invasion, especially when recombination is weak but nonzero. Allowing recombination itself to vary across generations, possibly jointly with mutation rate, would further generalize the problem to products of matrices with time-varying parameters, where invasion would depend on the joint distribution and covariance structure of transmission and linkage.

\noindent Empirically, the assumptions of strong selection, high mutation rates, large variance, and population-synchronized transmission are unlikely to apply universally. They may, however, be relevant in systems such as microbial populations experiencing episodic stress responses, transient hypermutability, error-prone polymerase activation, or environmentally induced changes in DNA repair. In such contexts, estimating temporal variation or tail behavior of mutation rates—rather than their means alone—may be informative, particularly when recombination reshuffles genetic backgrounds.

\noindent
Two main results emerge. First, in the absence of recombination ($R=0$), temporal variation in mutation rate affects invasion only through the geometric mean of transmission, lowering $\mathbb E[\log(1-u_{2,t})]$ relative to a deterministic rate with the same mean and thereby shrinking the region in which a rare modifier can invade. Second, when recombination is present ($R>0$), invasion need not vary monotonically with $R$: temporal stochasticity interacts with genetic mixing so that recombination, selection, and the resident mutation rate may enter the Lyapunov exponent nonlinearly, allowing the direction of selection on a rare modifier allele to change. Thus, without recombination variability has a purely quantitative effect, whereas with recombination it can qualitatively alter the dependence of invasion on other evolutionary parameters.

\newpage
\appendix
\section{Estimation of Invasion Regions from Lyapunov Exponents: Examples} \label{appendix:approximation}
\noindent At complete linkage ($R=0$), the per-generation invasion matrices are lower triangular for both ploidies. Consequently, the top Lyapunov exponent equals the larger of the two diagonal log–growth rates. Fix a resident mutation rate $u_1\in(0,1)$ and a rare modifier allele $M_2$ associated with i.i.d.\ mutation rates $u_{2,t}\in[0,L]$ ($L\le1$), with mean $u_2$ and variance $\sigma^2$. Then
\begin{equation}
\label{eq:lyap_clean}
\gamma(0)
\;=\;
-\log(1-u_1)
\;+\;
\max\!\Big\{
\mathbb E\big[\log(1-u_{2,t})\big],
\ \log \Lambda_{\mathrm{res}}
\Big\},
\end{equation}
where
\[
\Lambda_{\mathrm{res}}=
\begin{cases}
\dfrac{1-s}{\,1-u_1\,}, & \text{haploids},\\[6pt]
\dfrac{1-s(1+u_1)}{\,1-u_1\,}, & \text{diploids}.
\end{cases}
\]
\noindent The quantity $\Lambda_{\mathrm{res}}$ is the dominant growth factor of the resident genetic background in the absence of $M_2$. Under standard mutation–selection balance conditions ensuring the existence of a stable resident equilibrium—namely $u_1<s$ in haploids and $u_1<s/(1-s)$ in diploids—we have
\[
\log \Lambda_{\mathrm{res}} < \log(1-u_1),
\]
so that the resident background does not contribute to the asymptotic growth of the rare modifier allele. In this regime, \eqref{eq:lyap_clean} reduces to the unified invasion criterion
\begin{equation}
\label{eq:inv_unified}
\gamma(0)>0
\quad\Longleftrightarrow\quad
\mathbb E\!\big[\log(1-u_{2,t})\big]
\;>\;
\log(1-u_1).
\end{equation}
\noindent Criterion \eqref{eq:inv_unified} applies symmetrically to modifiers alleles that decrease
or increase mutation rates. A mutation–decreasing allele invades if its associated mutation process has a larger logarithmic mean survival than the resident rate $u_1$. Conversely, a mutation–increasing allele cannot invade unless temporal variation is sufficiently strong to raise $\mathbb E[\log(1-u_{2,t})]$ above $\log(1-u_1)$, despite having $\mathbb E[u_{2,t}]>u_1$. Thus invasion at complete linkage is governed by the logarithmic mean of the
mutation process, not by its arithmetic mean.

\noindent For any distribution $\mathcal L$ on $[0,L]$ with density $f(u)$, the expected log term is
\[
\mathbb E[\log(1-u_{2,t})]
\;=\;
\int_0^L \log(1-u)\,f(u)\,du.
\]
\noindent Once $\mathcal L$ is parameterized (e.g.\ by mean $u_2$ and variance $\sigma^2$), this integral can be evaluated either in closed form or by one–dimensional numerical integration. The invasion boundary at $R=0$ is then obtained by solving
\[
\mathbb E[\log(1-u_{2,t})]
\;=\;
\log(1-u_1),
\]
subject to the feasible support of $\mathcal L$.
\noindent Table~\ref{tab:lyap_G_summary_landscape} summarizes analytical and numerical expressions for $\mathbb E[\log(1-u_{2,t})]$ and the resulting invasion regions for the four distribution families considered.

\begin{landscape}
\begin{table}[p]
\centering
\small
\renewcommand{\arraystretch}{1.08}
\setlength{\tabcolsep}{5pt}

\vspace{6pt}
\resizebox{\linewidth}{!}{%
\begin{tabular}{>{\centering\arraybackslash}p{0.12\linewidth} >{\centering\arraybackslash}p{0.34\linewidth} >{\centering\arraybackslash}p{0.30\linewidth} >{\centering\arraybackslash}p{0.20\linewidth}}
\toprule
\textbf{Family on $[0,L]$} &
\textbf{Moment parameterization \& feasibility} &
\(\displaystyle \boldsymbol{\mathbb E[\log(1-u)]}\) &
\textbf{Invasion check at $R{=}0$} \\
\midrule

\textbf{Uniform} \(\ \mathrm{Unif}[a,b]\) &
\(u_2=\tfrac{a+b}{2},\ \sigma^2=\tfrac{(b-a)^2}{12}\).
Set \(a=u_2-\sqrt{3\,\sigma^2}\), \(b=u_2+\sqrt{3\,\sigma^2}\);
feasible if \(0\le a<b\le L\).&
\[
\frac{1}{b-a}\!\int_a^b\!\log(1-u)\,du
=\frac{H(b)-H(a)}{b-a},
\]
\(H(x)=(x-1)\log(1-x)-x\). &
Solve \(\mathbb E[\log(1-u)] = \log(1-u_1)\). \\[6pt]

\textbf{Beta} \(\ u=L X,\ X\sim\mathrm{Beta}(\alpha,\beta)\) &
\[
u_2=L\frac{\alpha}{\alpha+\beta},\quad
\sigma^2=L^2\frac{\alpha\beta}{(\alpha+\beta)^2(\alpha+\beta+1)}.
\]
Let \(m=u_2/L\) and \(v=\sigma^2/L^2\).
Then \(\alpha=m\,t,\ \beta=(1-m)\,t,\ t=\frac{m(1-m)}{v}-1\).
Feasible if \(0<u_2<L\) and \(0<\sigma^2<u_2(L-u_2)\). &
\[
\int_{0}^{1}\!\log(1-Lx)\,
\frac{x^{\alpha-1}(1-x)^{\beta-1}}{B(\alpha,\beta)}\,dx,
\]
Where \(B(\alpha,\beta)=\Gamma(\alpha)\Gamma(\beta)/\Gamma(\alpha+\beta)\). &
Map \((u_2,\sigma^2)\!\mapsto\!(\alpha,\beta)\), evaluate \(\mathbb E[\log(1-u)]\), and test \(>\log(1-u_1)\). \\[6pt]

\textbf{Truncated log–normal} \(\ \mathrm{LN}_{(0,L]}(\mu,\tau^2)\), \(\log u\sim\mathcal N(\mu,\tau^2)\) &
Any \((\mu,\tau^2)\) feasible on \((0,L]\).
Given \((u_2,\sigma^2)\), recover \((\mu,\tau^2)\) by matching truncated moments:
$u_2=\frac{\int_0^L u\,f_{\mathrm{LN}}(u;\mu,\tau^2)\,du}{\Phi\!\big(\tfrac{\log L-\mu}{\tau}\big)},$
$\sigma^2=\frac{\int_0^L u^2 f_{\mathrm{LN}}(u;\mu,\tau^2)\,du}{\Phi\!\big(\tfrac{\log L-\mu}{\tau}\big)}-u_2^2,$
where \(f_{\mathrm{LN}}(u)=\frac{1}{u\,\tau\sqrt{2\pi}}e^{-(\log u-\mu)^2/(2\tau^2)}\), \(\Phi=\) standard normal CDF. &
\[
\frac{1}{\Phi\!\big(\tfrac{\log L-\mu}{\tau}\big)}
\int_{0}^{L}\!\log(1-u)\,f_{\mathrm{LN}}(u;\mu,\tau^2)\,du.
\] &
Compute \(\mathbb E[\log(1-u)]\) (1D quadrature) and test \(>\log(1-u_1)\). \\[6pt]

\textbf{Truncated gamma} \(\ \mathrm{Ga}_{(0,L]}(\alpha,\beta)\) &
\(\alpha,\beta>0\) feasible. Truncated moments:
$u_2=\frac{1}{\beta}\frac{\gamma(\alpha+1,\beta L)}{\gamma(\alpha,\beta L)},$
$\sigma^2=\frac{1}{\beta^2}\frac{\gamma(\alpha+2,\beta L)}{\gamma(\alpha,\beta L)}-u_2^{\,2},$
$\gamma(a,z)=\int_0^z t^{a-1}e^{-t}\,dt.$ &
\[
\frac{1}{F_G(L;\alpha,\beta)}
\int_{0}^{L}\!\log(1-u)\,
\frac{\beta^\alpha}{\Gamma(\alpha)}u^{\alpha-1}e^{-\beta u}\,du,
\]
\(F_G(L;\alpha,\beta)=\gamma(\alpha,\beta L)/\Gamma(\alpha)\). &
Map \((u_2,\sigma^2)\!\mapsto\!(\alpha,\beta)\), evaluate \(\mathbb E[\log(1-u)]\), and test \(>\log(1-u_1)\). \\
\bottomrule
\end{tabular}%
} 

\caption{Summary of \(\mathbb E[\log(1-u_{2,t})]\) and the invasion test \(\mathbb E[\log(1-u_{2,t})]>\log(1-u_1)\) at \(R=0\) for four distribution families on \([0,L]\) (in simulations \(L=1\)). Means and variances refer to the distribution on \([0,L]\). Closed forms are available for some special cases (e.g., Beta with \(L{=}1\)); otherwise a single one–dimensional quadrature suffices.}
\label{tab:lyap_G_summary_landscape}
\end{table}
\end{landscape}

\section{Accuracy of Lyapunov Approximations} \label{appendix:lyap_acc}

\noindent
We evaluate the Lyapunov invasion criterion by comparison with stochastic recursion simulations ($5{,}000$ generations) under four distributions of mutation--rate supported on $[0,L]$: uniform, beta, truncated log--normal, and truncated gamma. The resident mutation rate is fixed at $u_1\approx0.048796$, selection is $s=0.2$, and recombination is absent ($R=0$).

\noindent With complete linkage, invasion is determined by the Lyapunov condition
\[
-\log(1-u_1)+\mathbb E\!\big[\log(1-u_{2,t})\big]=0,
\]
where $u_{2,t}$ is drawn i.i.d.\ across generations from the specified distribution. Let $u_2:=\mathbb E[u_{2,t}]$ denote the arithmetic mean of this distribution and $\sigma^2:=\operatorname{Var}(u_{2,t})$ its variance. For fixed $\sigma^2$, this equation defines a \emph{critical mean} $u_2^\ast$—a scalar mean value, not a realized mutation rate—that serves as a threshold: when the mean crosses $u_2^*$, the modifier allele’s long-run growth rate becomes negative, implying its eventual disappearance.

\noindent Uniqueness of $u_2^*$ holds conditional on fixed variance. Since $\log(1-x)$ is strictly decreasing on $[0,1)$, the mapping
\[
u_2 \longmapsto \mathbb E[\log(1-u_{2,t})]
\]
is strictly decreasing when $\sigma^2$ is held fixed. The Lyapunov equation therefore admits a unique root in the mean, and invasion occurs if 
\[
\mathbb E[\log(1-u_{2,t})]>\log(1-u_1).
\]
\noindent Thus invasion depends on the expected log of the non-mutation factor, not on whether $u_2$ lies above or below $u_1$.

\noindent
For each distribution family, we restrict $u_2$ to its feasible mean range given $\sigma^2$ and solve for $u_2^\ast$ using Brent’s method (tolerance $10^{-10}$). Distributional parameters are determined in closed form (uniform, beta) or by moment inversion (truncated log--normal, truncated gamma). Expectations are evaluated analytically when available (uniform and beta on $[0,1]$) and otherwise by one--dimensional adaptive quadrature, with endpoint splitting near $L=1$ to treat the logarithmic singularity.

\begin{figure}
\centering
\includegraphics[width=1.05\textwidth]{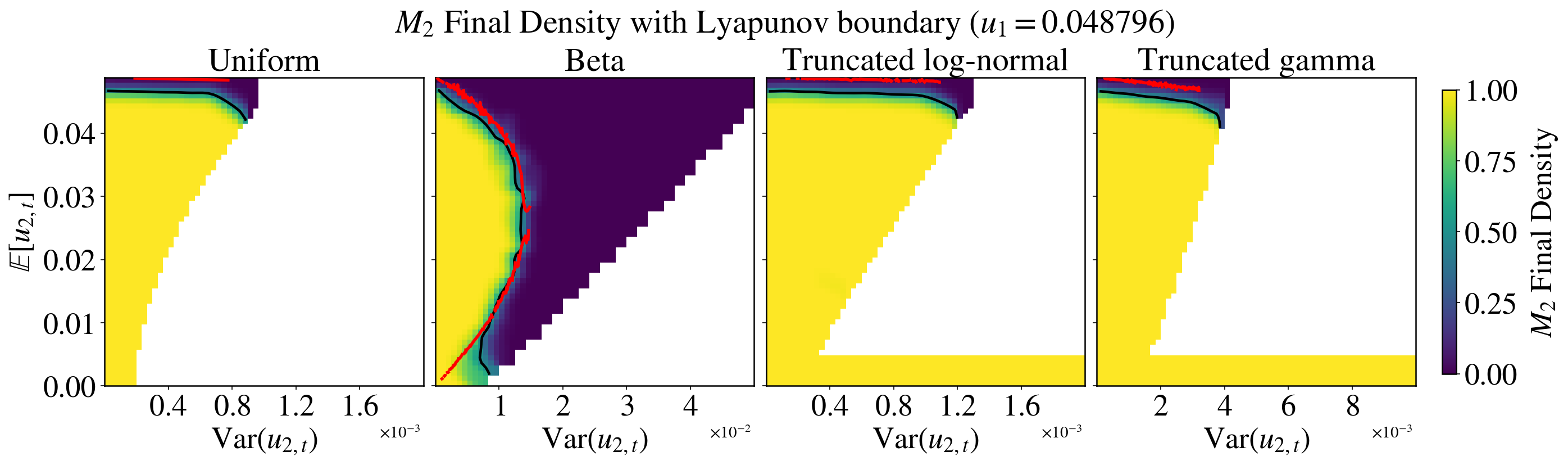}
\caption{\textbf{Fixation outcomes versus Lyapunov thresholds.} Each panel corresponds to one mutation–rate distribution. Settings: $5{,}000$ generations, $R=0$, $s=0.2$, $u_1\approx0.048796$. Colors indicate final $M_2$ density. The red curve shows the Lyapunov prediction $\gamma(u_2^\ast)=0$. Across all distributions, the predicted threshold closely tracks the observed transition between invasion and loss.}
\label{fig:lyapunov-critical-thresholds}
\end{figure}
\noindent
Figure~\ref{fig:lyapunov-critical-thresholds} shows close agreement between simulated fixation outcomes and the Lyapunov threshold. The Lyapunov prediction is mildly conservative: parameter combinations with $u_2<u_2^\ast(\sigma^2)$ reliably invade, whereas those with $u_2>u_2^\ast(\sigma^2)$ do not. Differences across distribution families arise from how, at fixed variance, probability mass is redistributed relative to $u_1$ and toward the upper boundary $1$, where $\log(1-u_{2,t})$ is highly sensitive.

\noindent
The beta family exhibits a non-monotone Lyapunov boundary in the $(u_2,\sigma^2)$ plane, with two critical variances for some fixed means. This does not imply multiple critical means. Conditional on fixed $\sigma^2$, the critical mean $u_2^\ast$ remains unique. The apparent multiplicity arises because varying $\sigma^2$ within the beta family necessarily alters higher moments and tail weight, inducing non-monotone dependence of $\mathbb E[\log(1-u_{2,t})]$ on variance. Thus the Lyapunov boundary may fold in $(u_2,\sigma^2)$-space while remaining single--valued in $u_2$ for each fixed $\sigma^2$.

\noindent
When most probability mass lies below $u_1$ and away from $1$—as for truncated log--normal distributions with thin upper tails, or for uniform distributions once their support includes $u_1$—the expected log multiplicative retention factor varies weakly with $\sigma^2$, yielding an approximately horizontal invasion boundary. In contrast, distributions that shift mass toward the right tail as variance increases (beta and truncated gamma) yield curved boundaries: increasing $\sigma^2$ lowers $\mathbb E[\log(1-u_{2,t})]$ and therefore lowers $u_2^\ast$.

\noindent
For example, at $\sigma^2\approx8\times10^{-4}$ the beta distribution yields $u_2^\ast\approx0.0484<u_1$, consistent with a modest right tail. The uniform family is infeasible for this variance because no support interval in $[0,1]$ can simultaneously realize the target variance and contain $u_1$. The truncated log--normal and truncated gamma distributions yield $u_2^\ast$ close to $u_1$ ($|u_2^\ast-u_1|\lesssim10^{-3}$), with the gamma slightly lower due to greater midrange weight. In all cases, reallocating probability mass from below $u_1$ toward the upper tail decreases $\mathbb E[\log(1-u_{2,t})]$ and shifts the invasion boundary downward, exactly as predicted by the Lyapunov criterion.

\section{Sensitivity of $\mathbb{E}[\log(1 - u_{2,t})]$ to Distributional Shape} \label{appendix: compare_derivations}
\noindent Fix the variance $\sigma^2\approx 0.00159$, support $(0,1)$, and resident mutation rate $u_1\approx 0.0954$. For each distributional family, define $u_2^*\in(0,1)$ as the largest mean mutation rate such that, under complete linkage ($R=0$),
\[
\gamma
\;=\;
-\log(1-u_1)+\mathbb{E}\!\big[\log(1-u_{2,t})\big]
\;=\;0.
\]
\noindent For $u_2<u_2^*$, the modifier allele experiences positive selection, whereas for $u_2>u_2^*$ it is selected against. Thus $u_2^*$ defines the invasion boundary in the case of fixed variance.

\noindent The function $\log(1-u)$ is strictly decreasing and strictly concave on $(0,1)$, with curvature increasing as $u\uparrow1$. Consequently, $\mathbb{E}[\log(1-u_{2,t})]$ is highly sensitive to probability mass near the upper boundary of the support. Small changes in the right tail of the distribution can therefore produce disproportionate changes in $\gamma$.

\noindent To make this dependence explicit, define
\[
\mathsf C(u_1):=\Pr(u_{2,t}\le u_1),
\qquad
\mathcal T(u_1):=\mathbb{E}\!\left[-\log(1-u_{2,t})\,\mathbf 1\{u_{2,t}>u_1\}\right],
\]
where $\mathbf 1\{\cdot\}$ is the indicator function, equal to $1$ when the condition inside the bracket is true and $0$ otherwise. The quantity $\mathsf C(u_1)$ measures the fraction of generations in which the modifier mutation rate does not exceed the resident rate, while $\mathcal T(u_1)$ measures the contribution to $-\log(1-u_{2,t})$ arising from realizations above $u_1$. Larger $\mathsf C(u_1)$ increases $\mathbb{E}[\log(1-u_{2,t})]$, whereas larger $\mathcal T(u_1)$ decreases it.

\noindent Table~\ref{tab:u2star_comparison} reports $u_2^*$ together with $\mathsf C(u_1)$ and $\mathcal T(u_1)$ for the four distributional families at the fixed $\sigma^2$ and $u_1$ considered here.

\begin{table}[t]
\centering
\renewcommand{\arraystretch}{1.12}
\setlength{\tabcolsep}{10pt}
\caption{Largest mean $u_2^*$ (at fixed $\sigma^2$) consistent with $\gamma=0$, together with the mass below $u_1$ and the upper–tail contribution.}
\begin{tabular}{lccc}
\toprule
\textbf{Distribution on $[0,1]$} & $\boldsymbol{u_2^*}$ & $\boldsymbol{\mathsf C(u_1)}$ & $\boldsymbol{\mathcal T(u_1)}$\\
\midrule
Uniform                & 0.09427879 & 0.507983 & 0.068280 \\
Beta                   & 0.09448161 & 0.559617 & 0.061754 \\
Truncated log-normal   & 0.09432226 & 0.591052 & 0.057932 \\
Truncated gamma        & 0.09427113 & 0.567150 & 0.060784 \\
\bottomrule
\end{tabular}
\label{tab:u2star_comparison}
\end{table}
\noindent The ordering of $u_2^*$ across distributions reflects a trade-off between mass below $u_1$ and the weight of the right tail. The beta distribution yields the largest $u_2^*$ by combining relatively high $\mathsf C(u_1)$ with a moderate upper–tail contribution. The truncated log-normal places the greatest mass below $u_1$ and has the smallest $\mathcal T(u_1)$, but gains little additional advantage because probability far below $u_1$ contributes weakly to $\mathbb{E}[\log(1-u_{2,t})]$. The uniform distribution allocates more mass above $u_1$, increasing $\mathcal T(u_1)$ and reducing $u_2^*$. The truncated gamma assigns comparatively more probability to intermediate-to-large mutation rates, further increasing $\mathcal T(u_1)$ and yielding the smallest $u_2^*$.

\noindent
Although the numerical differences in $u_2^*$ are small (on the order of $10^{-4}$), they are systematic. For fixed variance, shifting probability mass from the right tail toward values just below $u_1$ increases $\mathbb{E}[\log(1-u_{2,t})]$ and permits invasion at larger mean mutation rates. Conversely, shifting mass toward large mutation rates near the boundary $u\approx1$ lowers $\mathbb{E}[\log(1-u_{2,t})]$ and strengthens selection against the modifier. This applies symmetrically to mutation-decreasing modifiers ($u_2<u_1$), which may fail to invade if variability is sufficiently heavy-tailed, and to mutation-increasing modifiers ($u_2>u_1$), which cannot invade and are eliminated more rapidly as right-tail weight increases.

\section{Critical thresholds and nonlinear parameter dependence} \label{appendix:parameter_dependence}
For $R>0$, invasion is governed by the top Lyapunov exponent of a product of noncommuting random matrices, so the sign of $\gamma(\theta)$ may depend nonlinearly on any parameter $\theta$ that shapes modifier--background associations (e.g.\ $R$, $s$, or $u_1$). In particular, varying $\theta$ can change not only the magnitude but also the sign of $\gamma(\theta)$, and $\gamma(\theta)$ need not be monotone.

\noindent Roots of $\gamma(\theta)$ correspond to parameter values at which the long-run growth rate of a rare modifier vanishes. Because $\gamma(\theta)$ arises from a product of noncommuting random matrices, its dependence on $\theta$ need not be linear or monotone. Consequently, the equation $\gamma(\theta)=0$ may admit multiple solutions, each marking a qualitative change in invasion behavior.

\noindent
For any such parameter $\theta$ taking values in an interval $\Theta$, define the critical threshold
\[
\theta^\ast := \inf\{\,\theta\in\Theta:\gamma(\theta)=0\,\},
\]
the smallest value at which invasion changes direction. When $\gamma(\theta)$ is non-monotone, multiple roots of $\gamma(\theta)=0$ may exist; $\theta^\ast$ is the minimal root and therefore the minimal parameter change sufficient to reverse invasion.

\noindent Recombination provides a concrete illustration. Define
\[
R^\ast := \inf\{\,R\in[0,\tfrac12]:\gamma(R)=0\,\}.
\]
\noindent We estimate $R^\ast$ by evaluating $\gamma(R)$ on a fine grid using a common realization of $\{u_{2,t}\}$ across $R$, bracketing an interval $[R_L,R_U]$ with $\gamma(R_L)\gamma(R_U)<0$, and solving $\gamma(R)=0$ by bisection or Brent’s method. Convergence criteria are $|\gamma(R^\ast)|<10^{-5}$ or interval width $<10^{-3}$.
\begin{itemize}
\item \textit{Weak selection.} For $s=0.06$ and $u_{2,t}$ Beta-distributed with mean $u_2$ and variance $\sigma^2=0.25\,u_2(1-u_2)$, we obtain
\[
R^\ast \approx 0.27 \ \text{(haploid)},\qquad
R^\ast \approx 0.28 \ \text{(diploid)},
\]
consistent with the zero crossing near $R\in[0.25,0.30]$ in Fig.~\ref{fig:gamma_recombination}. In this regime, moderate recombination reverses invasion because it reduces the persistence of low-fitness genetic backgrounds disproportionately associated with $M_2$ during generations with elevated $u_{2,t}$. These high–$u_{2,t}$ episodes contribute strongly and negatively to the geometric mean growth rate; shortening their residence time increases $\gamma(R)$.

\noindent
At higher variance levels (e.g.\ $0.50$ or $0.75\,u_2(1-u_2)$), $\gamma(R)$ can become non-monotone in $R$. For small $R$, recombination primarily reduces the residence time of backgrounds generated during high–$u_{2,t}$ episodes, raising $\gamma(R)$. For larger $R$, recombination also reduces the persistence of high-fitness backgrounds formed during low–$u_{2,t}$ generations, when $M_2$ produces few new deleterious alleles and becomes concentrated on fitter backgrounds. The opposing effects can generate two zero crossings in $R$; by definition, $R^\ast$ denotes the smaller root. The existence and location of these roots depend on $(s,u_1)$ and on the full distribution of $u_{2,t}$, not merely its mean and variance.

\item \textit{Strong selection.} For $s=0.20$ and high variance $\sigma^2=0.95\,u_2(1-u_2)$, we find
\[
R^\ast \approx 0.46 \ \text{(haploid)},\qquad
R^\ast \approx 0.49 \ \text{(diploid)}.
\]
Here, reversal requires recombination $R$ close to $0.5$. Strong selection amplifies fitness differences among backgrounds, so the negative contribution of high–$u_{2,t}$ episodes dominates unless recombination is sufficiently frequent to continuously reshuffle the modifier onto higher-fitness backgrounds. In these cases, $\gamma(R)$ typically increases with $R$ for small $R$ and may decrease for larger $R$ once the loss of favorable low–$u_{2,t}$ backgrounds dominates.
\end{itemize}

\noindent The same procedure applies without modification to any parameter $\theta$. Thus, in regimes where analytic invasion conditions are unavailable, one can (i) detect nonlinear dependence of invasion on $\theta$ by inspecting the shape of $\gamma(\theta)$ and (ii) compute context-specific critical values $\theta^\ast$ that separate invasion from loss.

\section{Supplementary Information}

\subsection*{Constant Mutation Rate ($u_2$)}
\label{sec:invasion_deterministic_unified}
\noindent We consider the case in which the mutation rate associated with modifier allele $M_2$ is constant through time.  This corresponds to the special case $u_{2,t}\equiv u_2$ for all $t$, with $\mathbb{E}[u_{2,t}]=u_2$ and $\mathrm{Var}(u_{2,t})=0$.  
\noindent Near the resident mutation--selection equilibrium
\[
\hat{\mathbf x}=(\hat x_1,\hat x_2,0,0)^\top,
\]
maintained by modifier allele $M_1$ with mutation rate $u_1$, we introduce a rare modifier allele $M_2$ with mutation rate $u_2$.  Linearizing the two--locus recursion at $\hat{\mathbf x}$ yields
\[
\bm v_{t+1}=\mathbf F\,\bm v_t,
\]
where $\bm v_t=(x_{3,t},x_{4,t})^\top$ collects the rare $M_2$--bearing haplotypes and $\mathbf F$ is a nonnegative $2\times2$ matrix.  By the Perron--Frobenius theorem \cite{frobenius1908_1909}, $\mathbf F$ has a unique real leading eigenvalue $\lambda_+=\rho(\mathbf F)>0$, which governs the asymptotic growth of $M_2$.  The modifier invades if $\lambda_+>1$.

\noindent Let $p(\lambda)=\lambda^2-\tau\lambda+\delta$ denote the characteristic polynomial of $\mathbf F$, where $\tau=\mathrm{tr}(\mathbf F)$ and $\delta=\det(\mathbf F)$.  Since
\[
p(\lambda)=(\lambda-\lambda_-)(\lambda-\lambda_+),
\qquad 0<\lambda_-<\lambda_+
\]
at the resident polymorphic equilibrium, it follows that
\[
p(1)=(1-\lambda_-)(1-\lambda_+).
\]
Because $\lambda_-<1$ under mutation--selection balance, the condition $\lambda_+>1$ is equivalent to $p(1)<0$.  Hence invasion can be determined by evaluating $p(1)=1-\tau+\delta$ at the resident equilibrium.

\noindent Direct calculation gives
\[
\text{haploids: }\quad
p(1)=\frac{(s-u_1)(u_2-u_1)}{(1-u_1)^2},
\qquad
\text{diploids: }\quad
p(1)=\frac{\big[s(1+u_1)-u_1\big](u_2-u_1)}{(1-u_1)^2}.
\]
\noindent For haploids, existence of a polymorphic mutation--selection balance requires $0<u_1<s$, ensuring $s-u_1>0$; for additive diploids, the corresponding condition is $0<u_1<\tfrac{s}{1-s}$, implying $s(1+u_1)-u_1>0$.  Under these admissible parameter ranges, the sign of $p(1)$ is therefore determined solely by the difference $(u_2-u_1)$, yielding
\begin{equation}
\label{eq:deterministic_reduction_principle}
u_2<u_1 \;\Longrightarrow\; \lambda_+>1,
\qquad
u_2>u_1 \;\Longrightarrow\; \lambda_+<1.
\end{equation}
\noindent Thus, a modifier allele that reduces the mutation rate invades, whereas an allele that increases the mutation rate cannot invade. This is the classical \emph{Reduction Principle}.

\paragraph{Effect of recombination.}
\noindent Although recombination does not affect the invasion \emph{criterion}---since $p(1)=1-\tau+\delta$ is independent of $R$---it does affect the \emph{rate} of invasion through its influence on the leading eigenvalue $\lambda_+$.  \noindent Writing
\[
\lambda_+=\tfrac{1}{2}\bigl(\tau+\sqrt{\tau^2-4\delta}\bigr)
\]
and differentiating the characteristic equation $p(\lambda)=0$ implicitly with respect to $R$ yields
\[
\frac{d\lambda_+}{dR}
=\frac{\tau'(R)\lambda_+(R)-\delta'(R)}{2\lambda_+(R)-\tau(R)}.
\]
\noindent Since $p(1)$ is invariant under changes in $R$, we have $-\tau'(R)+\delta'(R)=0$, hence $\delta'(R)=\tau'(R)$ and
\begin{equation}
\label{eq:pull_to_one_general}
\frac{d\lambda_+}{dR}
=\frac{\tau'(R)\big(\lambda_+(R)-1\big)}{2\lambda_+(R)-\tau(R)}.
\end{equation}
\noindent For a $2\times2$ matrix, $2\lambda_+-\tau=\lambda_+-\lambda_->0$, so the sign of $d\lambda_+/dR$ is determined by $\tau'(R)$ and $\lambda_+-1$.

\noindent It remains to compute $\tau'(R)$.
\begin{itemize}
    \item \emph{Haploids.} From \eqref{eq:F_hap},
    \[
    \tau_{\mathrm{hap}}(R)=\frac{1}{1-u_1}\!\left[(1-u_2)\!\left(1-\frac{u_1R(1-s)}{s(1-u_1)}\right)+(1-s)-(1-u_2)\frac{(s-u_1)R(1-s)}{s(1-u_1)}\right],
    \]
    so
    \begin{equation}\label{eq:tauprime_hap}
    \tau'_{\mathrm{hap}}(R)=-\,\frac{(1-u_2)(1-s)}{s(1-u_1)^2}\;<\;0
    \qquad\text{for }u_1,u_2\in(0,1),\ s\in(0,1).
    \end{equation}
    
    \item \textit{Diploids.} From \eqref{eq:F_dip}, the trace is
    \[
    \begin{aligned}
      \tau_{\mathrm{dip}}(R)
      &= \frac{1}{(1-u_1)s} \biggl[
          (1-u_2)\bigl(s - u_1 R(1-s)\bigr) \\
      &\quad
          + (1-s)\bigl((1-R) + u_2 R\bigr)\bigl(s - u_1(1-s)\bigr)
          + (1-2s)u_1
        \biggr].
    \end{aligned}
    \]
    hence
    \begin{equation}\label{eq:tauprime_dip}
    \tau'_{\mathrm{dip}}(R)=-\,\frac{(1-u_2)(1-s)}{\,1-u_1\,}\;<\;0
    \qquad\text{for }u_1,u_2\in(0,1),\ s\in(0,1).
    \end{equation}
\end{itemize}

\noindent In both haploids and diploids, explicit differentiation of the trace yields
\[
\tau'(R)<0
\qquad
\text{for }u_1,u_2\in(0,1),\ s\in(0,1).
\]
\noindent Substituting into \eqref{eq:pull_to_one_general} gives
\begin{equation}
\label{eq:pull_to_one_sign}
\operatorname{sign}\!\Big(\tfrac{d\lambda_+}{dR}\Big)
=\operatorname{sign}\!\big(1-\lambda_+(R)\big).
\end{equation}
\noindent Thus, recombination monotonically shifts $\lambda_+$ toward unity: it slows invasion when $\lambda_+>1$ and accelerates loss when $\lambda_+<1$.  When $u_2=u_1$, $\lambda_+(R)\equiv1$ for all $R$, and recombination has no effect, consistent with classical results
\citep{karlin1974random,altenberg1984,feldman1986evolutionary,liberman1986modifiers,altenberg2017unified}.

\paragraph{Effects of selection and mutation parameters.} When $u_{2,t}\equiv u_2$, the parameters $s$ and $u_1$ do not alter the \emph{direction} of selection on the modifier, which is fixed by the sign of $u_2-u_1$, but they do modulate the \emph{magnitude} of the growth factor $\lambda_+$.  Since
\[
p(1)=(1-\lambda_-)(1-\lambda_+),
\]
we have
\[
\lambda_+-1=-\,\frac{p(1)}{1-\lambda_-}.
\]
\noindent At the resident mutation--selection balance $0<\lambda_-<1$, so $s$ and $u_1$ affect invasion only through the magnitude of this ratio, not its sign.

\noindent Substituting the explicit expressions for $p(1)$ yields
\begin{equation}
\label{eq:lambda_plus_minus_one_decomposition}
\lambda_+-1
=
-\,\frac{u_2-u_1}{(1-u_1)^2}\,
\frac{c(s,u_1)}{1-\lambda_-},
\qquad
c(s,u_1)=
\begin{cases}
s-u_1, & \text{haploids},\\
s(1+u_1)-u_1, & \text{diploids}.
\end{cases}
\end{equation}
\noindent Thus, $u_2-u_1$ determines the sign of invasion, while $s$ and $u_1$ enter only as scaling factors.  As $s$ approaches the lower bound for maintaining a mutation--selection balance, $c(s,u_1)\to0$ and $\lambda_+\to1$, reflecting the disappearance of indirect selection as $a$ alleles cease to be efficiently removed.  Away from this boundary, increasing $s$ or $u_1$ increases $|\lambda_+-1|$, strengthening invasion when $u_2<u_1$ and accelerating loss when $u_2>u_1$.

\noindent In summary, with a constant mutation rate the evolution of a modifier allele is entirely governed by the Reduction Principle.  Selection, mutation, and recombination modulate only the \emph{rate} of invasion or loss, not its direction.  This clean separation between qualitative and quantitative effects provides the natural baseline against which the consequences of stochastic transmission can be evaluated.

\newpage
\section*{Code and supplementary material}

All code and supplementary material used for the analyses in this paper are available at
\url{https://github.com/ElisaHeinrich/Evo_Stochastic_Transmission_Mut_Modifiers}.

\bibliography{report.bib} 
\bibliographystyle{abbrv}
\end{document}